\documentclass[a4paper, amsfonts, amssymb, amsmath, reprint, showkeys, nofootinbib, twoside]{revtex4-1}
\usepackage[english]{babel}
\usepackage[utf8]{inputenc}
\usepackage{ulem}

\usepackage[most]{tcolorbox}

\usepackage{xr}
\makeatletter
\newcommand*{\addFileDependency}[1]{
  \typeout{(#1)}
  \@addtofilelist{#1}
  \IfFileExists{#1}{}{\typeout{No file #1.}}
}
\makeatother

\newcommand*{\myexternaldocument}[1]{
    \externaldocument{#1}
    \addFileDependency{#1.tex}
    \addFileDependency{#1.aux}
}
\myexternaldocument{appendix}

\usepackage[colorinlistoftodos, color=green!40, prependcaption]{todonotes}
\usepackage{amsthm}
\usepackage{mathtools}
\usepackage{physics}
\usepackage{xcolor}
\usepackage{graphicx}
\usepackage[left=23mm,right=13mm,top=35mm,columnsep=15pt]{geometry} 
\usepackage{adjustbox}
\usepackage{placeins}
\usepackage[T1]{fontenc}
\usepackage{lipsum}
\usepackage{csquotes}
\usepackage{comment}

\usepackage[pdftex, pdftitle={Article}, pdfauthor={Author}]{hyperref} 
\begin{document}
\renewcommand{\figurename}{Fig.}
\title{From One to Two Dimensions: Magnetic Phases in Weakly Coupled Spin Ladders}

\author{Mateo Cárdenes Wuttig}
    \email[Correspondence email address: ]{mateo.cardeneswuttig@yale.edu}%
    \affiliation{Department of Physics, Columbia University, New York, NY 10027, United States}
    \affiliation{Department of Applied Physics, Yale University, New
Haven, Connecticut 06520, United States}

\author{Andrew J. Millis}
    \affiliation{Center for Computational Quantum Physics, The Flatiron Institute, 162 5th Avenue, New York, NY 10010, United States}
    \affiliation{Department of Physics, Columbia University, New York, NY 10027, United States}

\date{\today} 

\begin{abstract}
A large variety of materials can be approximately described by means of spin-1/2 Heisenberg ladders. Here, the Density Matrix Renormalization Group (DMRG) algorithm together with a previously established numerical self-consistent mean-field approximation is used to investigate the magnetic properties of spin ladders coupled in a second dimension. The full ground state phase diagram including spin-gapped, antiferromagnetic, ferrimagnetic and fully polarized phases is presented as a function of interladder and intraladder coupling and magnetic field.
Measurement of the dependence of magnetization on applied magnetic field is shown to enable location of a material on the phase diagram and determination of the Hamiltonian parameters. These results provide a practical route toward identifying and characterizing magnetic materials composed of coupled spin ladders.
\end{abstract}

\maketitle

\section{Introduction \label{sec:Intro}}
A spin ladder (see fig. \ref{fig:system}a) is a one-dimensional system consisting of two chains ($j=1,2$ in fig. \ref{fig:system}a) with an interaction $J_\parallel$ coupling spins  along the chains  and also a 'rung' ($J_\perp$) interaction coupling adjacent spins in different  chains. If each site hosts a spin $1/2$ and the interactions are antiferromagnetic and SU(2) symmetric (Heisenberg), the ground state has no long-ranged order, has a gap to spin excitations, and may be viewed as a quantum superposition of singlet pairs. Spin ladders are conceptually interesting as examples of spin liquid states. It has been theoretically proposed\cite{doi:10.1126/science.273.5281.1515,Dagotto_1999} that adding mobile carriers to a spin ladder will produce strong superconducting correlations persisting to high temperature. 

Many materials, including some copper-oxide perovskite-derived compounds \cite{PhysRevB.49.8901, doi:10.1021/acs.chemmater.2c02939,PhysRevB.95.104428}, and molecular materials\cite{PhysRevLett.86.5168,D1TC01576A,PhysRevB.41.1657,PhysRevB.103.205131,sous2021}, can be approximately described as coupled spin-1/2 Heisenberg ladders, and an intriguing possibility is that doping these compounds may lead to novel superconductivity. However, in all of these cases the question of deviations from one-dimensionality is an important issue. The development of experimental protocols for determining if a given material may be viewed in terms of weakly coupled spin ladders and is therefore likely to become superconducting on doping is of interest.

The magnetic properties and excitation spectrum of a single spin ladder are well understood\cite{doi:10.1126/science.271.5249.618,PhysRevLett.73.882,PhysRevLett.73.886}. An important step towards understanding the effects of interladder coupling was taken by Giamarchi and collaborators\cite{PhysRevLett.101.137207,PhysRevB.83.054407,bouillot2012staticsdynamicsweaklycoupled}. Motivated by measurements on the organic compound bis(piperidinium)tetrabromocuprate(II) (BPCB), Giamarchi et al. introduced a formalism combining a numerically exact treatment of the properties of a single ladder with a mean field treatment of the interladder coupling. These and previous authors explored the behavior as a function of applied magnetic field and temperature showing that coupled spin ladders can have two phases: a singlet phase and a magnetically ordered phase. Building on this work, we present a comprehensive treatment of the interladder coupling between adjacent spin ladders for small and large intraladder couplings in a magnetic field, including two different coupling geometries relevant to different materials of current experimental interest. We investigate the critical interladder coupling and present several diagnostics to determine the Hamiltonian parameters from magnetization data. We use the Density Matrix Renormalization Group (DMRG) algorithm in the Matrix Product State (MPS) formulation as the numerically exact treatment of a single ladder. We investigate both a system where all rungs are coupled, similar to BPCB\cite{PhysRevB.41.1657, PhysRevB.83.054407}, and one where only every second rung is coupled, which is realized, for example, in terphenyl\cite{sous2021}. Because in many cases measurements of magnetization ($M$) as a function of applied field ($H$) are feasible whereas direct probes of spin correlations and antiferromagnetic ordering are not, we focus on obtaining and interpreting results for $M(H)$.

We note that the  spin-flop transition in classical antiferromagnets \cite{10.1080/00018739700101558,Kumar_2013} presents similar issues. In low-dimensional classical antiferromagnets above a lower critical field, a state with non-zero magnetization develops. With increasing magnetic field, this state saturates to full polarization above an upper critical field. The field dependence of the magnetization in this case helps pin down the values of the different exchange constants of classical antiferromagnets.

Our work is structured as follows. In section ~\ref{sec:methods}, we introduce the Hamiltonian of coupled spin ladders and explain the different magnetic phases. In section ~\ref{sec:results}, we present the zero-temperature phase diagram as a function of magnetic field and coupling strength between ladders. Next, we present magnetization data for different interladder and intraladder couplings and show how the critical fields, midpoint magnetization, and the slope of $M(H)$ can be used to determine the system parameters. We determine the critical interladder coupling to induce staggered order at zero field and explain how the quantitative phase diagram depends on coupling parameters and geometry.

\section{Model and Methods\label{sec:methods}}
\subsection{Hamiltonian}
\begin{figure}[htbp!]
 \centering
 \begin{adjustbox}{center}
   \includegraphics[width=0.75\columnwidth]{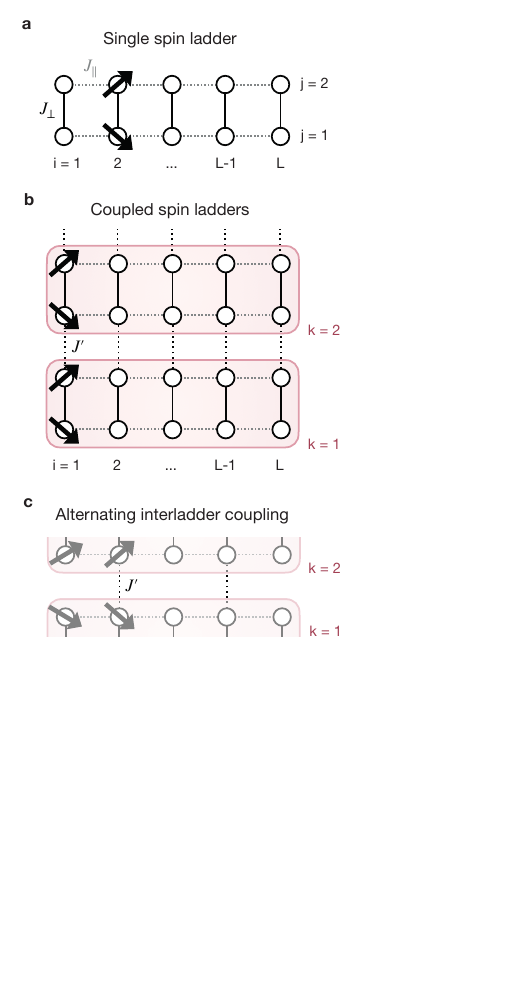}
 \end{adjustbox}
 \caption{Depiction of spin ladder models. (a) Single spin ladder of length $L$ with coupling $J_{\perp}$ on rungs $i$ and $J_{\parallel}$ on legs $j$. (b) Coupled spin ladders with interladder coupling $J'$ are effectively a two-dimensional system with $J_{\parallel}$ along $x$-direction and alternating $J_{\perp}$ and $J'$ in $y$-direction. Each ladder $k$ is visualized by a red box. (c) Alternating interladder coupling $J'$ between two ladders $k=1,2$ on every second rung.}
 \label{fig:system}
\end{figure}

In this work, we study a system of parallel spin ladders, with nearest neighbor ladders coupled by a Heisenberg interaction. An isolated ladder, depicted in fig. \ref{fig:system}(a), is a one-dimensional system with a unit cell containing two sites, each holding a $S=1/2$ local moment described by the three conventional Heisenberg spin operators $\vec{S}$ coupled by a Hamiltonian
\begin{equation}
	H_{\text{spin ladder}}= H_{\perp}+ H_{\parallel}+H_{\text{mag}}
\end{equation}
comprising a \textit{rung} Hamiltonian
\begin{equation}
	H_{\perp} = J_{\perp}\sum_{i=1}^L \vec{S}_{i,1} \cdot \vec{S}_{i,2} \,,
\end{equation}
a \textit{chain} Hamiltonian
\begin{equation}
    	H_{\parallel} = \gamma J_\perp  \sum_{i=1}^L \sum_{j=1}^2 \vec{S}_{i,j} \cdot \vec{S}_{i+1,j} \,,
\end{equation}
and a coupling to a uniform external magnetic field
\begin{equation}
H_{\text{mag}}=- h^{z} J_\perp \sum_{i=1}^L\sum_{j=1}^2 S^z_{i,j} \,.
\end{equation}
A ladder of length $L$ thus has $2L$ sites denoted as $(i,j)$, where $i=1,...,L$ is the position on each of the two legs $j=1,2$. We choose to measure energies in units of $J_\perp$ and have defined the dimensionless intraladder coupling $\gamma=J_{\parallel}/ J_\perp$ and the dimensionless magnetic field strength $h_z$.

The Hamiltonian of $m$ parallel spin ladders coupled in the second dimension is 
\begin{equation}\label{eq:coupledspinladders}
	H_{m \, \text{spin ladders}}= \sum_{k=1}^m H_{\text{spin ladder},k} + H'
\end{equation}
We consider two different geometries for interladder coupling: One, motivated by the BPCB compound, in which every site in one ladder is coupled to one site in another ladder, and one, motivated by the recently synthesized terphenyl compound\cite{sous2021}, in which only every other rung is coupled in the second dimension. The Hamiltonian for BPCB\cite{PhysRevB.41.1657} and related materials, see fig. \ref{fig:system}(b), is:
\begin{equation}
    H' = J' J_\perp \sum_{k=1}^{m-1} \sum_{i=1}^L \vec{S}_{i,2,k} \cdot \vec{S}_{i,1,k+1} \,,
\end{equation}
where we have defined a dimensionless interladder coupling $J'$ and $k=1,...,m$ labeling different ladders.
The Hamiltonian for terphenyl\cite{sous2021}, see fig. \ref{fig:system}(c), is
\begin{equation}
    H'_{\text{terphenyl}} = J' J_\perp \sum_{k=1}^{m-1} \sum_{\rm i\, even} \vec{S}_{i,2,k} \cdot \vec{S}_{i,1,k+1} \,.
\end{equation}
The absence of some of the interladder couplings in terphenyl leads to  features that will be discussed below.
We note that in real materials, there will also be an interplane coupling $J_{=}$ connecting two-dimensional layers of coupled spin ladders. Usually, $J_{=}$ is much smaller than all other energy scales, so it is irrelevant to the ground state properties we investigate in this work and is not considered here.

\subsection{Phases and Observables}
We introduce two important observables for a single spin ladder: the uniform $(m^z)$ and staggered $(m^x_a)$ magnetization, defined as
\begin{equation}
    m^z = \frac{1}{2L} \sum_{i=1}^L\sum_{j=1}^2 S^z_{i,j}
\end{equation}
and
\begin{equation}
    m^x_a = \frac{1}{2L} \sum_{i=1}^L\sum_{j=1}^2 (-1)^{i+j} S^{x}_{i,j} \,,
\end{equation}
respectively. 

The model of coupled ladders may be viewed as a square-lattice Heisenberg model with different couplings along different bonds. Depending on the relative values of the parameters, we may expect antiferromagnetically ordered phases ($m^x_a\neq0$), spin-gapped phases ($m^z=m^x_a=0$) and, for sufficiently large magnetic fields, a fully spin-polarized phase ($m^z=1/2;~m^x_a=0$).
At $J'=0$ we have a decoupled set of spin ladders. Except at $\gamma=\infty$\cite{PhysRevB.47.3196}, an isolated ladder at zero field ($h=0$) has a ground state with exponentially decaying correlations and a gap $\Delta$ to spin excitations. For all $\gamma$, $\Delta=c_\gamma J_\perp$ with $c_\gamma=1-\mathcal{O}(\gamma)$ at small $\gamma$ and $c_{\gamma} \approx \frac{1}{2}$ as $\gamma\rightarrow\infty$.\cite{PhysRevB.69.092408}.
 
\begin{figure}[htbp!]
 \centering
 \begin{adjustbox}{center}
   \includegraphics[width=\columnwidth]{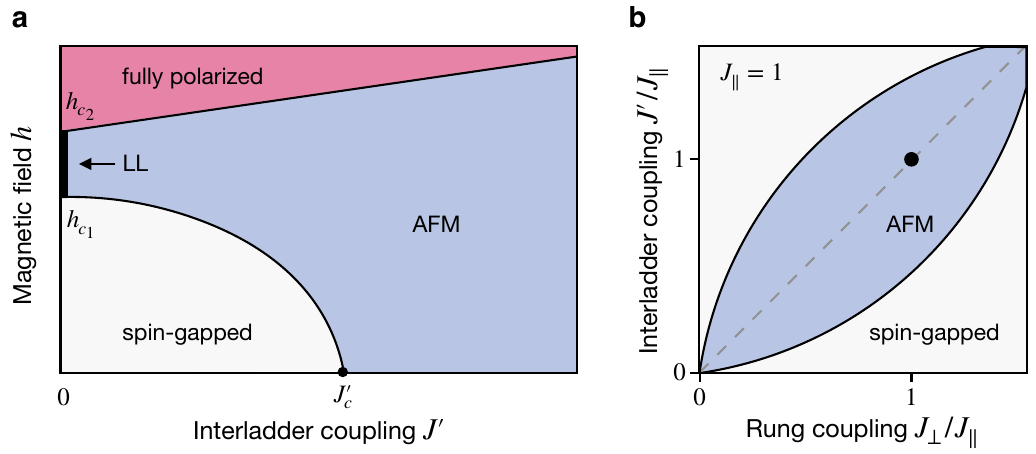}
 \end{adjustbox}
 \caption{(a) Schematic phase diagram of coupled spin ladders as a function of interladder coupling $J'$ (horizontal) and magnetic field $h$ (vertical) sketched for generic intraladder coupling $\gamma$. At small magnetic field below $h_{c_1}$ and interladder coupling $J' < J'_c$, the system is in a spin-gapped, decoupled ladder phase (gray). At fields larger than an upper critical field $h_{c_2}$, the system is fully spin polarized (pink). At intermediate fields the system possesses antiferromagnetic order (purple) except at $J'=0$ where one has decoupled Luttinger liquids (thick solid line). (b) Zero-field ($h = 0$) phase diagram as a function of rung coupling (horizontal) and interladder coupling (vertical) relative to leg coupling $J_{\parallel}$ shows symmetry between $J_{\perp}$ and $J'$. Black dot indicates square lattice Heisenberg model.}
 \label{fig:phasediagram2}
\end{figure}

The presence of the gap means that this state is stable to small perturbations. For uncoupled ladders, at a critical magnetic field $h_{c_1} \sim \Delta$, there occurs a transition to an intermediate Luttinger liquid (LL) state with staggered (antiferromagnetic) power-law correlations. At any nonzero interladder coupling $J'$, the LL state acquires long-ranged antiferromagnetic order at temperature $T=0$\cite{PhysRevB.83.054407}. At zero magnetic field, a transition to an antiferromagnetic state occurs when $J'$ is increased above a critical value  $J'_c\sim \Delta$. The transitions may be simply understood from the excitation spectrum of the isolated ladder at zero field. The first excited state is a triplet ($S=1$) of energy $\Delta$; applying a positive field $h$ in the $z$-direction lowers the energy of the $S^z=1$ component of the triplet, while the interladder coupling $J'$ also lowers the energy of the magnon mode. When the energy of this state vanishes, the state reconstructs\cite{PhysRevB.83.054407}. Above a larger field $h_{c_2}$, the system becomes fully polarized, so the magnetization saturates, and the antiferromagnetic order parameter becomes zero\cite{PhysRevB.83.054407}. This sequence of phases is presented in fig. \ref{fig:phasediagram2}(a). 

The above considerations pertain to a generic, not too small $\gamma$. The special case of $\gamma = 0$ (decoupled rungs) requires additional discussion. In the BPCB case (fig. \ref{fig:system}(b)) at $J'\neq 0$, but $\gamma=0$, we have decoupled spin chains with a coupling that alternates between $J'$ and $J_\perp$ along the chain. Except for the special case $J'=J_\perp$\cite{betheansatz,doi:10.1126/science.271.5249.618}, each chain has a spin gap, implying that also for small $\gamma$ the ground state is generically spin gapped at all $J'$\cite{PhysRevB.59.11384}. However, for larger, "generic" $\gamma$, shown in fig. \ref{fig:phasediagram2}(a), we expect that for $J' > J'_c\sim\Delta$, an antiferromagnetic ground state occurs. 

In this paper, we characterize the different transitions and provide quantitative estimates that enable extraction of system parameters from measurements of the magnetization as a function of field. We are interested in observables that will help in locating a system on the phase diagram in fig. \ref{fig:phasediagram2}(b). The $m^z(h)$ curve reveals three important field scales: 
\begin{itemize}
    \item $h_{c_1}$: the lower critical field at which $m^z$ first becomes nonzero, 
    \item $h_{c_2}$: the upper critical field at which the spins become fully polarized, and
    \item $h_m$, the magnetic field at which $m^z=1/4$. 
\end{itemize}
The upper critical field $h_{c_2}$ is analytically known, see \cite{PhysRevB.59.11398, PhysRevB.55.3046, Chaboussant1998} and supplemental section \ref{sec:appendix_hc2}.\cite{supp} Restoring physical units 
\begin{equation}\label{eq:hc2}
h_{c_2}= 2J_{\parallel} + J_\perp + J'
\end{equation}
The lower critical field $h_{c_1}$ does not, in general, have an analytical dependence on model parameters. If the interladder coupling is large enough, $h_{c_1}=0$; if the interladder coupling is very small, $h_{c_1}=J_\perp\left(1-c \gamma\right)$ with $c$ varying smoothly from $1$ at $\gamma=0$\cite{PhysRevB.83.054407} to $1/2$ for $\gamma\gg1$\cite{PhysRevB.69.092408} and $h_{c_1}$ decreases with increasing $J'$. Information available from the midpoint value will be discussed below.

The magnetization curve has an interesting scaling for $h\approx h_{c_1}$. For an isolated ladder, $m^z(h)$ shows a square root scaling\cite{PhysRevB.83.054407, PhysRevB.59.11398, PhysRevB.50.258}
\begin{equation}\label{eq:critscaling_sqrt}
    m^z \sim \sqrt{h^z - h_{c_1}} \quad \text{and} \quad m^z \sim 1/2 - \sqrt{h_{c_2} - h^z} 
\end{equation}
while in the ordered phase for coupled spin ladders $m^z(h)$ scales linearly
\begin{equation}\label{eq:critscaling_lin}
    m^z \sim (h^z - h_{c_1}) \quad \text{and} \quad m^z \sim 1/2 - (h_{c_2} - h^z) \,.
\end{equation}
with a crossover to the square root scaling visible if $J'$ is small enough.
The scaling forms hold because the triplet excitations can be mapped to repulsive hard-core bosons, implying that  the transitions belong to the Bose-Einstein condensate universality class\cite{PhysRevB.83.054407, PhysRevB.59.11398}.

\subsection{Numerical Mean-Field Theory}
In most systems, we expect a comparably weak interladder coupling $J' \ll J_{\perp}, J_{\parallel}$, which is why we expect the mean-field decoupling between ladders first introduced by Giamarchi et al.\cite{PhysRevB.83.054407} to accurately describe the effect of the interladder interactions. We decouple the interactions between two adjacent rungs $i$ on ladders $k$ and $k+1$ and treat the interladder coupling as an external mean-field in 
the following mean-field Hamiltonian
\begin{equation}
    H_{\text{MF spin ladder}} = H_{\text{spin ladder}} + J' J_{\perp}  H_{\text{MF}} 
\end{equation}
\begin{equation}
    H_{\text{MF}} = m^z \sum_{i=1}^L\sum_{j=1}^2 S^z_{i,j} + 2 m^x_a \sum_{i=1}^L\sum_{j=1}^2 (-1)^{i+j} S^x_{i,j}
\end{equation}
where $m^z$ and $m^x_a$ are determined self-consistently. We use $m^z_{\text{init}} = 0$ and $m^x_{a,\text{init}} = 0.5$ for all calculations and define a cutoff of $\epsilon = 10^{-3}$ for $m^z$ and $m^x_a$ to determine convergence between two iterations. 
For a system with alternating interladder coupling, see fig. \ref{fig:system}(c), we apply the mean-field terms only on every second rung.

\subsection{DMRG}
We use the Density Matrix Renormalization Group (DMRG) algorithm\cite{PhysRevLett.69.2863, RevModPhys.77.259} in Matrix Product State (MPS) form\cite{dmrg_uli} to compute the zero-temperature ground states of our Hamiltonians. This method is particularly well suited for (quasi)-one-dimensional systems. All our calculations were performed using the ITensor library\cite{itensor}. In the supplemental section, we verify the convergence of our calculation with respect to system size and present finite-size studies of a single spin ladder and of our mean-field calculations.\cite{supp} We use a typical bond dimension of $m=500$ and system sizes of $L=30$ to reach convergence.

\section{Results \label{sec:results}}
\subsection{Magnetic Phase Diagram}
In this section, we show in detail how an applied uniform magnetic field can be used to locate an experimental system of interest on the phase diagram and to extract the critical behavior, focusing on a terphenyl system. We show that the phase diagram does not change qualitatively for a spin ladder system in which every rung is coupled.

\begin{figure}[htbp!]
 \centering
 \begin{adjustbox}{center}
   \includegraphics[width=\columnwidth]{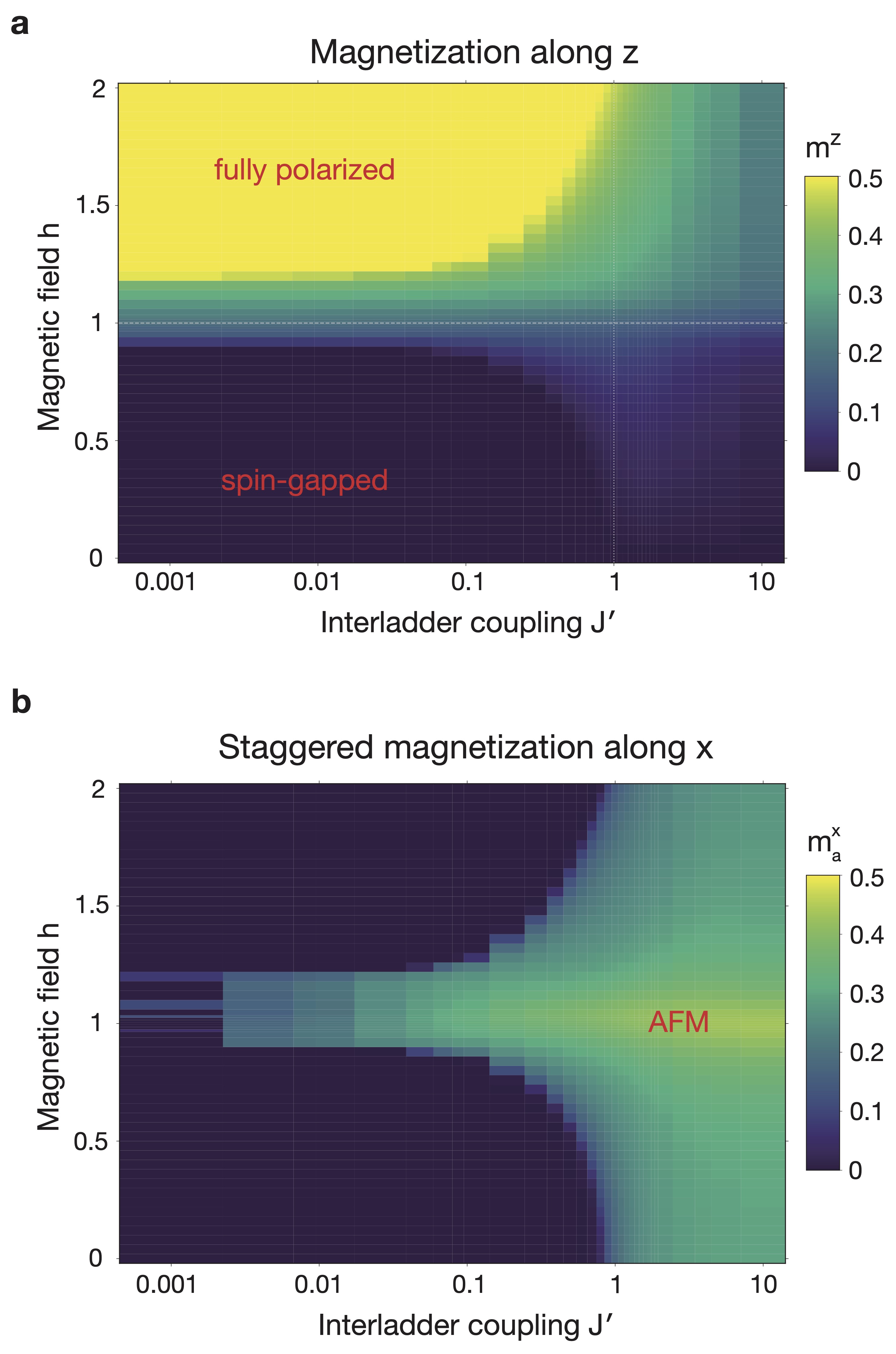}
 \end{adjustbox}
 \caption{Ground state phase diagram as a function of interladder coupling $J'$ (horizontal axis) and magnetic field $h$ (vertical axis) obtained with DMRG and self-consistent numerical mean-field theory for a spin ladder with $\gamma = 0.1$. (a) Magnetization along $z$ reveals the unpolarized spin-gapped phase and fully polarized phase. (b) Staggered magnetization along $x$ shows the boundaries of the antiferromagnetic phase (AFM).}
 \label{fig:fig2}
\end{figure}

In fig. \ref{fig:fig2}, we present the phase diagram of a spin ladder of length $L=30$ with mean-field couplings $J'$ on every second rung and a small intraladder coupling $\gamma = 0.1$. We perform calculations for different magnetic fields $h \in [0,2]$ (vertical axis) and interladder couplings $J' \in [0.001, 10]$ (horizontal axis).
Based on the magnetization along $z$ and staggered magnetization along $x$, see figs. \ref{fig:fig2}(a) and (b), respectively, we can observe several different phases:\cite{PhysRevB.83.054407} 
\begin{itemize}
    \item \textit{Spin-gapped phase}: At small interladder coupling $J'$, we find a gapped spin liquid phase with a singlet ground state, where the average magnetization is $m^z = 0$ for magnetic fields smaller than $h_{c_1}$. The gap defined by $h_{c_1}$ decreases with increasing  $J'$ and eventually vanishes above $J'_c$.
    \item \textit{Gapless phase}: For larger magnetic fields, we encounter a gapless phase between $h_{c_1}$ and $h_{c_2}$ with increasing $m^z = 0$ to $m^z = 1/2$. 
    \item \textit{Fully polarized phase}: For $h > h_{c_2}$, we have a fully polarized ground state characterized by $m^z = 1/2$, meaning that all spins are parallel to $h^z$. 
    \item \textit{Antiferromagnetic phase}: When we increase the interladder coupling $J' > J'_c$ at $h < h_{c_1}$ ($h > h_{c_2}$), we encounter a competition between the spin-gapped (fully polarized) and antiferromagnetic phase. With increasing magnetic field $h >0$, the antiferromagnetic phase also occurs for smaller $J'$, but only in the gapless phase for $h_{c_1} < h < h_{c_2}$. There is always a $h_{c_2}$ above which the ordered phase is destroyed.
\end{itemize}

In the antiferromagnetic phase, we find that $m^x_a \ll 1/2$ at zero field, and $m^x_a \rightarrow 1/2$ for magnetic fields $h_{c_1} < h < h_{c_2}$, and for $J' \rightarrow \infty$, where the system is in the gapless phase. 
Increasing the field $h\gg1$ induces a non-zero magnetization. The phases and their boundaries do not change when we allow coupling between all rungs, see fig. \ref{fig:fig_phase_diagram_gamma_0}, only the strength of the staggered order increases. The dotted horizontal (vertical) lines at $h = 1$ ($J' = 1$) indicate when the magnetic field (interladder coupling) is equal to the rung coupling $J_{\perp} = 1$. Note that the staggered order at $J'=0.001$ is incorrect due to finite size effects.

\subsection{Magnetization Curves}
In this section, we present and discuss our computed magnetization data showing how measured magnetization curves may be used to infer the parameters of the theory. Another technique for extracting the coupling parameters is via nuclear magnetic resonance (NMR) experiments measuring the nuclear spin-lattice relaxation rate $T_1^{-1}$ as a function of the magnetic field and comparing it to DMRG results of both $m^z$ and $m^x_a$\cite{PhysRevB.83.054407,PhysRevLett.101.137207}. We characterize the curves by $h_{c_1}, h_{c_2}, h_m$ and the slope (differential susceptibility) $\alpha=\dv{m^z(h)}{h}$, where $\alpha(h \rightarrow 0)$ is the uniform susceptibility. We separately consider the cases of weak and strong intraladder coupling.

\subsubsection{Weak Intraladder Coupling}
\begin{figure*}[htbp!]
 \centering
 \begin{adjustbox}{center}
   \includegraphics[width=2.05\columnwidth]{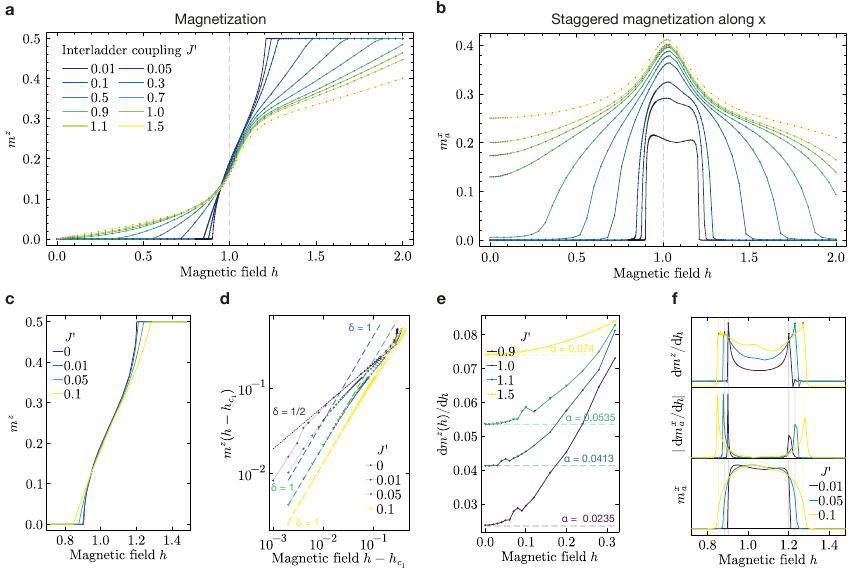}
 \end{adjustbox}
 \caption{Magnetization data for a spin ladder with small intraladder coupling $\gamma = 0.1$ and mean-field coupling on every second rung. (a) Uniform magnetization along $z$ plotted against applied field $h$ for interladder couplings $J'$ shown in panel legend. For $J^\prime \leq 0.5$ the data show a spin gap ($m^z=0$ for a range of fields). By $J'=0.9$ the model is in a magnetic state with a non-zero susceptibility even as $h\rightarrow 0$. (b) Staggered magnetization along $x$ shows the boundaries of the antiferromagnetic phase. For $J' > J'_c$, the system has a nonzero staggered order even at zero magnetic field. (c) Expanded view of $m^z(h)$ from panel (a), showing weak dependence on $J'$ for small $J'$, with deviations becoming apparent for $J'=\gamma$. $J' = 0$ is data from a spin ladder without mean-field couplings.
(d) Magnetization plotted against difference of field from critical value $h_{c_1}$. A region of linear scaling ($\delta = 1$) followed by a crossover to the square root scaling ($\delta = 1/2$) expected for an isolated chain is evident. (e) For larger $J' > J'_c$, the slope $\alpha = \dv{m^z(h)}{h}$ at $h\rightarrow 0$ increases with increasing $J'$. (f) The staggered order (lower plot) and its derivative (middle plot) can be inferred from the derivative of the magnetization (upper plot) and the inflection points (vertical lines). Normalized data in arbitrary units.}
 \label{fig:fig3}
\end{figure*}
A spin ladder with weak intraladder coupling $\gamma \ll 1$ has spin-singlets that are tightly bound on the individual rungs. 
In fig. \ref{fig:fig3}(a), we present the magnetization curves for a spin ladder with $\gamma = 0.1$ of length $L=30$ coupled in mean-field with different interladder couplings $J'$ on every second rung. The decrease of the lower critical field $h_{c_1}$ with $J'$ is evident, and we see that the critical $J'_c$ for closing the spin gap is $J'_c \approx 1$. 
We observe that, for the small $\gamma$ case considered here, the magnetization curve is approximately symmetric about its midpoint for small interladder coupling, see fig. \ref{fig:fig3}(c); the asymmetry becomes more pronounced for larger $J'$. The field $h_m$ at which we observe the midpoint magnetization is defined as $m^z(h_m) = 1/4$. For small $\gamma$, we can estimate\cite{PhysRevB.83.054407} $h_m$ to be
\begin{equation}\label{eq:hm_small_gamma}
    h_m = \frac{h_{c_1} + h_{c_2}}{2} = 1 + \gamma/2 \,,
\end{equation}
allowing us to determine the coupling $J_{\perp} = 1$ in absolute units. Here, $h_m \approx 1.05$ for a spin ladder with a small intraladder coupling $\gamma = 0.1$. We observe that $h_m$ is very insensitive to the interladder coupling: even for large $J' = 1.5$, we find that $h_m \approx 1.12$. The midpoint $h_m$ is close to the inflection point of the magnetization curve, allowing us to identify $h_m$ even if $h_{c_2}$ is not accessible. 
While $h_{c_1} = J_{\perp} - J_{\parallel}$ only works well for small $J'$, we find that 
\begin{equation}\label{eq:hc2}
    h_{c_2} \approx 1 + 2\gamma + J'
\end{equation} still holds for larger interladder couplings; see, for example, the case of $J' = 0.7$. In the supplemental section, we prove that this result for $h_{c_2}$ is exact for all geometries of coupled spin ladders considered in this work.\cite{supp}
This means that we can determine $J_{\perp}$ from $h_m$ and $J_{\perp} + 2J_{\parallel} + J'$ from $h_{c_2}$.
We do not find good agreement for $h_m$ in eq. \ref{eq:hm_small_gamma} when we use $h_{c_2}$ from eq. \ref{eq:hc2}.

Panel (b) of fig. \ref{fig:fig3} shows the amplitude $m^x_a$ of the staggered magnetization. As expected on general grounds, $m^x_a\neq 0$ when $0<m^z<1/2$. Hence, we can use $m^z(h)$ to infer $m^x_a(h)$, since the phase boundaries are nearly identical. In particular, at small $J'$, the rise in $m^x_a(h)$ is very sharp. The magnetic field $h$ at which the staggered order becomes maximal decreases slightly with stronger interladder coupling $J'$. We find that for small $J'$ and fixed $\gamma$, the system is gapped and one can use the spin gap, defined by the lower critical field $h_{c_1}$, to determine $J'$. In fig. \ref{fig:fig3}(c), we plot magnetization curves around the gapless phase for small $J'$, including the magnetization curve of a single spin ladder without mean-field couplings, denoted as $J' = 0$. We use a system of $L=100$ for $J'=0$. The curves for $J' = 0$ and $J' = 0.01$ seem identical in the main part of the gapless phase with indistinguishable shapes, but they differ close to the critical field. 

Panel (d) of fig. \ref{fig:fig3} shows a log-log plot of $m^z(h)$ versus the field measured from the lower critical field for small $J'$. We see a range of linear scaling (dashed line, $\delta = 1$) that decreases as $J'$ decreases, crossing over to the square root scaling expected for an isolated chain. The crossover between these two scalings appears at $h - h_{c_1} \sim J'$. Note that the scaling of $J'=0$ deviates from the square root behavior denoted as $\delta = 1/2$ (dotted line) for $h - h_{c_1} < 10^{-2}$ due to finite size effects discussed in supplemental fig. \ref{fig:fig_finite_size_04}(d).\cite{supp}

\begin{figure}[htbp!]
 \centering
 \begin{adjustbox}{center}
   \includegraphics[width=\columnwidth]{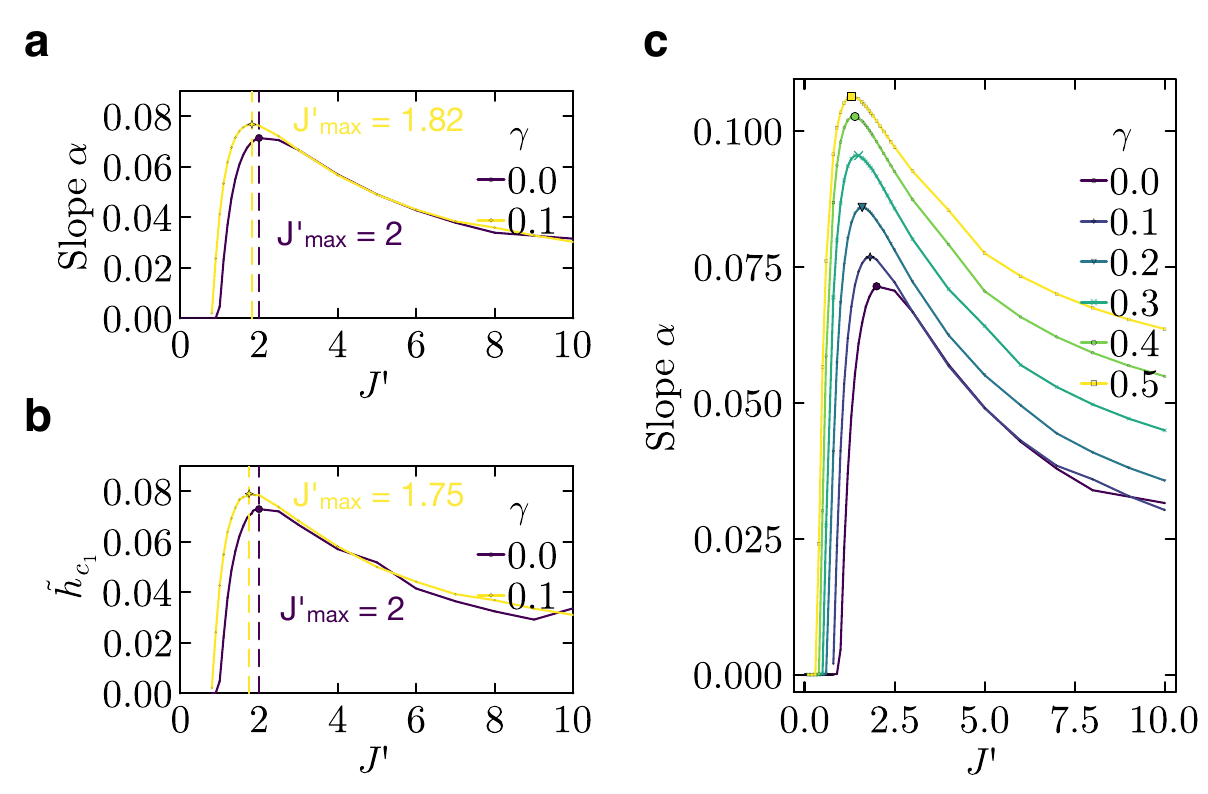}
 \end{adjustbox}
 \caption{(a) Slope $\alpha(h \rightarrow 0)$ close to zero field for $\gamma = 0$ and $\gamma = 0.1$ as a function of interladder coupling $J'$ shows a maximum at $J'_{max}$ (large symbols). (b) Critical field $\tilde{h}_{c_1}$ above which the differential susceptibility deviates significantly from the linear response value versus $J'$ with $J'_{max,\tilde{h}}$. (c) Slope $\alpha(h \rightarrow 0)$ as a function of $J'$ shows increasing maximal slope $\alpha_{max}$ (large symbols) at decreasing $J'_{max}$ with increasing $\gamma$.}
 \label{fig:fig3b}
\end{figure}

For larger $J'$, we encounter a gapless system with staggered magnetic order even at zero field, see fig. \ref{fig:fig3}(b). Here, one can use the slope $\alpha$ close to $h = 0$ (i.e., the linear response susceptibility) to identify $J'$. In fig. \ref{fig:fig3}(e), we see that $\alpha$ increases with increasing $J'$. In the following, we show that the linear response susceptibility is a sensitive measure of the interladder coupling, while it does not depend strongly on the intraladder coupling; see fig. \ref{fig:fig3b}.

Figure \ref{fig:fig3}(f) shows the staggered magnetization along $x$ (lower plot) and its slope (middle plot) versus the magnetization along $z$ (upper plot) in normalized units. Although $h_{c_1}' \approx h_{c_1}$, we find that there exists a small nonzero $m^x_a$ in the gapped and fully polarized phase for small $J'$ close to the critical field. The first derivative of the magnetization $\dv{m^z(h)}{h}$ (upper plot) can be used to infer the antiferromagnetic order, both through the onset of the magnetization ($h_{c_1}$ and $h_{c_2}$), see also fig. \ref{fig:fig3}(c), and via the inflection points, i.e., the maxima of $\dv{m^z(h)}{h}$ visualized as vertical lines (upper plot) and $\dv{m^x_a(h)}{h}$ (middle plot). The antiferromagnetic order parameter becomes nonzero at the same magnetic field at which a nonzero magnetization appears.

Systems with larger interladder coupling have a ground state with staggered order and $m^z > 0$ for any $h\neq 0$, see fig. \ref{fig:fig3}(a), leading to $h_{c_1} = 0$. In this case it is useful to define a new critical field $\tilde{h}_{c_1}$ above which the slope begins to deviate appreciably from the linear response value. In practice, we determine this threshold for non-linearity as $\alpha(h = \tilde{h}_{c_1})/\alpha(h \rightarrow 0) = 1 + p$, where we use $p = 2\%$
In fig. \ref{fig:fig3b}, we compare the slope $\alpha(h \rightarrow 0)$ close to zero field, see panel (a), to $\tilde{h}_{c_1}$ in panel (b) for two different values of $\gamma$.
Both $\alpha(J')$ and $\tilde{h}_{c_1}(J')$ show the same dependence on $J'$, including a maximum value at $J'_{max,\alpha}$ or $J'_{max,\tilde{h}}$, and monotonically increasing or decreasing behavior for $J' < J'_{max}$ and $J' > J'_{max}$, respectively. For the decoupled rung limit ($\gamma = 0)$, we find that $J'_{max,\alpha} = J'_{max,\tilde{h}}$, while at $\gamma = 0.1$, we find $J'_{max,\alpha} \approx J'_{max,\tilde{h}}$. 
Hence, we may only use the slope of $m^z(h)$ to characterize $J'$, which is easier since $\alpha$ can be calculated more precisely and at smaller magnetic fields than $\tilde{h}_{c_1}$. Additionally, $\alpha$ has the natural interpretation of how close a system with staggered order is to its phase boundary. 

In fig. \ref{fig:fig3b}(c), we present the slope as a function of interladder coupling for a range of intraladder couplings $\gamma$. As expected, there is no unique solution to solve $\alpha(\gamma,J')$ without knowing $m^x_a\vert_{h=0}$ for fixed $\gamma$ or $J'$ or having an estimate of $\gamma$ or $J'$, for example, from ab initio density functional theory (DFT) calculations, allowing us to first fix $J'$ from $m^x_a$ and then $\gamma$, or vice versa.

\subsubsection{Strong Intraladder Coupling}
\begin{figure*}[htbp!]
 \centering
 \begin{adjustbox}{center}
   \includegraphics[width=2.05\columnwidth]{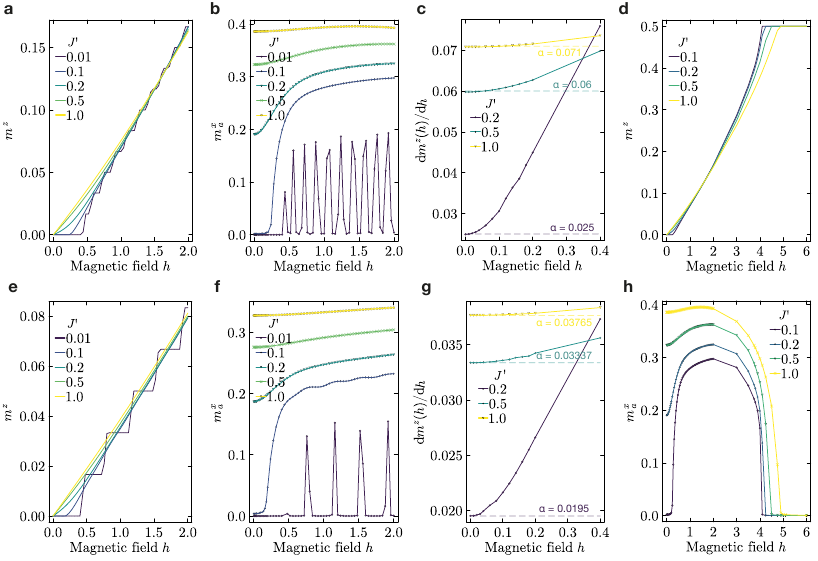}
 \end{adjustbox}
 \caption{Magnetization data for spin ladder with larger intraladder couplings (a)-(c) $\gamma = 1.5$ and (e)-(g) $\gamma = 3$ with mean-fields on every other rung. (a) Magnetization along $z$ for different interladder couplings between $J' = 0.01$ and $J' = 1$ indicates the presence of a spin gap for $J' \leq J'_c \approx 0.1$. (b) Staggered magnetization along $x$, with AFM order at zero field for $J' > J'_c$. Note that curve for $J' = 0.01$ shows peaks due to finite size effects discussed in the supplemental section. (c) The slope $\alpha(h)$ of the magnetization increases with $J'$. (d) Magnetization along $z$ for a larger range of magnetic fields up to $h = 6$. (e)-(g) $m^z$, $m^x_a$, and $\alpha(h)$, respectively, for $\gamma = 3$. As in panels (a) and (b), finite size effects are visible for the two smallest $J^\prime$-values. (h) $m^x_a$ for a larger range of magnetic fields for $\gamma = 1.5$ shows boundaries of staggered order, which is destroyed above $h_{c_2}$.}
 \label{fig:fig4}
\end{figure*}

Next, we discuss the magnetic behavior of spin ladders with larger intraladder couplings. In fig. \ref{fig:fig4}(a) and (e), we present the magnetization data for a spin ladder of length $L=30$ coupled in mean-field on every second rung with larger intraladder couplings of $\gamma = 1.5$ and $\gamma = 3$, respectively. For small $J' \leq 0.1$, the system is gapped with no staggered order at zero field, see fig. \ref{fig:fig4}(b) and (f). The spin gap of a single spin ladder decreases with increasing $\gamma$, see fig. \ref{fig:figures_SI_single_spin_ladder}, although $h_{c_1} \approx J_{\perp}/2$.\cite{PhysRevB.69.092408} Hence, the presence of a spin gap indicates that $J' < J'_c$. In this case, we can measure $h_{c_1}$, and thus determine $J'$ for fixed $\gamma$. For $J' > J'_c$, the only parameter directly accessible is the slope, which gives us the absolute value of $J'$. We observe that the slope $\alpha(h \rightarrow 0)$ becomes larger with increasing $J'$, even up to large interladder couplings at $J' = 1$, which is visualized in fig. \ref{fig:fig4}(c) and (g). 
In certain cases, we will also observe $h_m$, allowing us to identify the dominating energy scale $\gamma$ as 
\begin{equation}\label{eq:hm_large_gamma}
    h_m \approx 2\gamma
\end{equation} 
In fig. \ref{fig:fig4}(d) and (h), we present data for the magnetization and staggered magnetization, respectively, for a larger range of magnetic fields. Again, we observe that $m^z$ is insensitive to $J'$ in the central part of the magnetization curve. In comparison with fig. \ref{fig:fig3}, we find that the overall shape of $m^z(h)$ and $m^x_a(h)$ is independent of $\gamma$. Additionally, we find that the exact solution for $h_{c_2}$ allows us to extract a linear combination of the coupling parameters. There is no remaining staggered order at large magnetic field $h > h_{c_2}$, and the system is fully polarized. 

Especially for large intraladder couplings, we may only be able to access fields below $h_m$. If we know the absolute value of the magnetization, we can use $\alpha$ to infer how close we are to the phase boundary, since $\alpha \rightarrow 0$ for $J' \rightarrow J'_c$, or by comparing it to our estimates, see also fig. \ref{fig:fig3b}(c). Additionally, we can extrapolate $h_m$ and $h_{c_2}$ by solving $1/2 = m^z(h_{c_2}) = \alpha (h_{c_2} - h_{c_1})$ for $h_{c_2}$. This method is not highly accurate, but it provides us with a rough estimate of the model parameters. 

It may be possible that we have no estimate of $m^z$ in absolute units, and we cannot observe any features except $\alpha$. In this case, we can use $\tilde{h}_{c_1}$, the threshold for non-linearity, and define a new field scale $\tilde{h}_{c_2}$ as the field at which we observe the first inflection point of $\alpha$. The latter field scale is estimated as $\tilde{h}_{c_2} = \text{arg max}_{h}\left(\dv{\alpha(h)}{h}\right)$.
We can determine an absolute energy scale of the slope $\tilde{h}_{c_1} \approx \alpha$. For small $J' < 0.5$, we find 
that
\begin{equation}\label{eq:tilde_hc1}
    \tilde{h}_{c_1} \approx J'/10
\end{equation}
and
\begin{equation}\label{eq:tilde_hc2}
    \tilde{h}_{c_2} \approx J'/2
\end{equation}
These  fields scales are explored in detail in the supplemental section in fig. \ref{fig:fig4SI}.\cite{supp}

\subsection{Critical Interladder Coupling and Ladder Geometry}
\begin{figure*}[htbp!]
 \centering
 \begin{adjustbox}{center}
   \includegraphics[width=1.5\columnwidth]{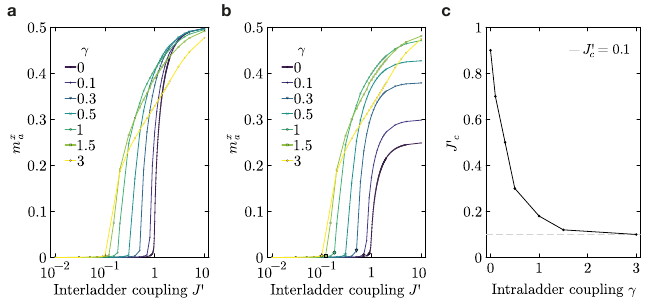}
 \end{adjustbox}
 \caption{Staggered magnetization along $x$ at zero field $h=0$ versus interladder coupling $J'$ for a spin ladder with different intraladder couplings $\gamma$. Data for system with mean-fields on (a) every rung and (b) every second rung. The boundaries of the antiferromagnetic phases increase with increasing $\gamma$. (c) Critical interladder coupling decreases from $J'_c \approx 0.9$ in the single rung limit $\gamma = 0$ to $J'_c \approx 0.1$ for large intraladder coupling. Data corresponds to large symbols in (b).}
 \label{fig:fig5}
\end{figure*}

In fig. \ref{fig:fig5}(a), we present the zero-field ($h = 0$) staggered magnetization for different $\gamma$ as a function of interladder coupling $J'$ for a system where every rung is coupled. There is no staggered order, even for large intraladder coupling, below the critical interladder coupling $J'_c \approx 0.1$ for $\gamma > 1$. This means that the system is always gapped for small $J'$. Figure \ref{fig:fig5}(b) shows the same data but for a system in which only every second rung is coupled. 
For $\gamma > 1$, both curves of $m^x_a(J')$ in fig. \ref{fig:fig5}(a) and (b) are nearly identical. Both geometries have a ground state without staggered order ($m^x_a(J' < J'_c) \approx 0$) below $J'_c$.
The only difference between a system where every rung is coupled and one where every second rung is coupled, see fig. \ref{fig:fig5}(a) and (b), respectively, is the behavior of $m^x_a$ for large $J'$ and small $\gamma < 1$. In the former case, $m^x_a \rightarrow 1/2$, independent of $\gamma$, while in the latter system, the magnitude of $m^x_a$ decreases to $m^x_a \rightarrow 1/4$ in the limit of $\gamma \rightarrow 0$, and $m^x_a < 1/2$ for $\gamma < 1$. Figure \ref{fig:fig5}(c) shows that $J'_c$ (corresponding to large symbols in fig. \ref{fig:fig5}(b)) decreases to $J'_c \approx 0.1$ for large $\gamma$. $J'_c(\gamma)$ is independent of the number of coupled rungs. Note that the result of $J_c \approx 0.9$ in the decoupled rung limit for $\gamma = 0$ may be misleading since the ground state at $J' = 1$ is gapless, as discussed previously. Thus, this curve should only be used to infer the data for small $\gamma > 0$.
We expect that $J'_c$ scales with the spin gap, which goes to zero for $\gamma \rightarrow \infty$. Note that we use units of $J_{\perp} = 1$, in which the gap is $\Delta \approx J_{\perp}/2$ for larger $\gamma$\cite{PhysRevB.69.092408}. Therefore, we can confirm $J_c/J_{\perp} \approx 0.1$ for $\gamma \gg 1$\cite{D1TC01576A}, while $J_c/J_{\perp} \gg 0.1$ for $\gamma < 1$.
Hence, we can extrapolate our phase diagram presented in fig. \ref{fig:fig2}(b) to larger values of $\gamma$, where fig. \ref{fig:fig5}(b) corresponds to a horizontal slice along $h=0$.

With the magnetization data for small and large intraladder couplings in figs. \ref{fig:fig3} and \ref{fig:fig4}, respectively, we show that the schematic phase diagram in fig. \ref{fig:phasediagram2} is independent of $\gamma$. For a single spin ladder with small $\gamma \lesssim 0.3$ and without interladder coupling ($J' = 0$), the gapless phase spans a width of $3\gamma$ from $h_{c_1}$ to $h_{c_2}$. For larger $\gamma$, the width of the gapless region increases less strongly with $\gamma$. The width of the gapless region also increases with $J'$, see fig. \ref{fig:fig3} for small $\gamma$, until the system becomes gapless at $J'_c \approx 0.1$ for $\gamma \gg 1$ and $J'_c < 0.9$ for $\gamma \ll 1$, as shown in fig. \ref{fig:fig5}. Hence, the parameter region of the gapless phase in fig. \ref{fig:fig2} can be estimated from the coupling parameters. 

\begin{figure}[htbp!]
 \centering
 \begin{adjustbox}{center}
   \includegraphics[width=\columnwidth]{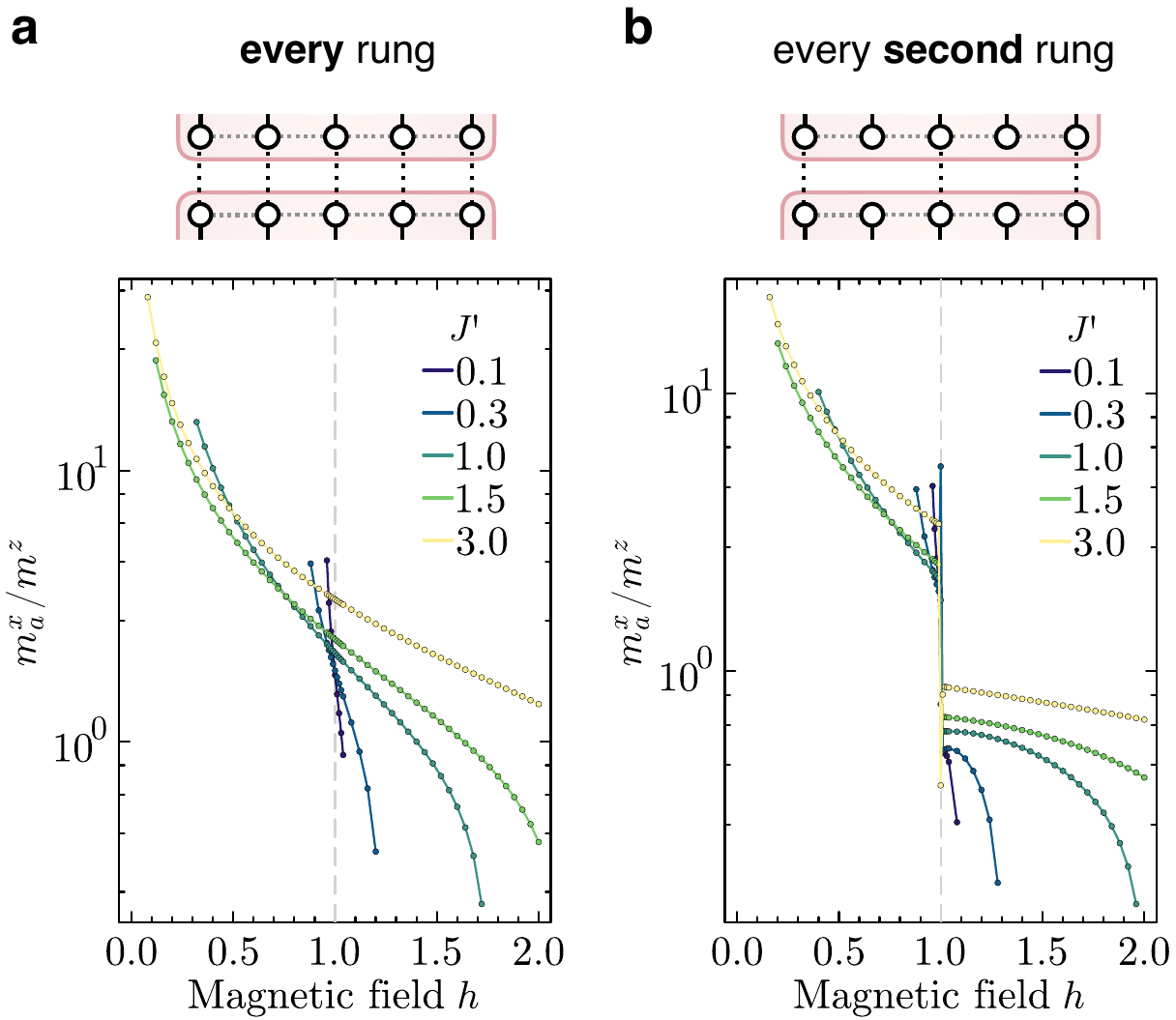}
 \end{adjustbox}
 \caption{Ratio of staggered magnetization along $x$ ($m^x_a$) to magnetization along $z$ ($m^z$) versus magnetic field. All calculations are performed with $\gamma = 0$. We can clearly observe the difference between (a), a system where every rung is coupled, which shows a smooth transition for $h<J_{\perp}$ and $h>J_{\perp}$, and (b), a system where every second rung is coupled, where $m^x_a/m^z$ jumps for $h>J_{\perp}$ in units of $J_{\perp} = 1$ (dashed vertical line).}
 \label{fig:fig6}
\end{figure}
 
The critical interladder coupling $J'_c$ is the same whether every rung or every second rung is coupled with $J'$, and the only difference between these two cases is the magnitude of the staggered order for small $\gamma$. In fig. \ref{fig:fig6}, we show the ratio of staggered magnetization to magnetization in the decoupled rung limit of $\gamma = 0$ for a system where every rung is coupled (panel (a)) compared to one where every second rung is coupled, see panel (b). Note that we only plot data in the staggered phase with $0.49 > m^z > 0.01$ and $m^x_a > 0.01$. We can clearly observe the difference around $h = 1$: while $m^x_a/m^z (h)$ is smoothly decreasing with increasing field around $h \approx 1$ for a spin ladder where every rung is coupled, we see that there is a large jump from $h<J_{\perp}$ to $h>J_{\perp}$ in panel (b) for a system where every second rung is coupled. Thus, the ratio of $m^x_a/m^z$ can be used as a sensitive measure to determine whether all rungs of a spin ladder or only a subset of them are coupled with an interladder coupling $J'$ by comparing values for magnetic fields above and below the energy scale set by the rung coupling.

All of our results are calculated at zero temperature. We know that the system becomes disordered at high temperatures, destroying both AFM and fully polarized phases in the high temperature limit\cite{sous2021}. The magnetization increases at low but non-zero temperature when the system is either in the spin-gapped or gapless phase at zero temperature, and decreases again at higher temperatures. The crossover temperature between an antiferromagnetic state and decoupled spin ladders is estimated to be $T_c \approx J'$\cite{PhysRevB.83.054407}.

\section{Discussion and outlook \label{sec:discussion}}
In this work, we provide detailed calculations for a previously introduced mean-field theory for coupled spin ladders by Giamarchi et al.\cite{PhysRevB.83.054407} for systems with arbitrary interladder and intraladder couplings. Additionally, we show that this approach can also be used for systems such as terphenyl\cite{sous2021} where only every second rung is coupled in the second dimension. We present a quantitative zero-temperature phase diagram to localize the magnetic phases, including the unpolarized paramagnetic phase, gapless phase, fully polarized phase, and antiferromagnetic phase, as a function of intraladder coupling, interladder coupling, and magnetic field. We describe how the critical magnetic fields depend on the system parameters and provide formulas to relate these characteristic values. Our data show that we can use the critical fields and the slope of the magnetization to determine the interladder coupling based on experimental $M(H)$ data. Additionally, we determine the critical interladder coupling $J'_c$ to induce staggered order at zero magnetic field as a function of intraladder coupling and show that there is no antiferromagnetic order below $J'_c \approx 0.1$, even for systems with large intraladder coupling. We argue that the qualitative phase diagram does not depend on the system parameters; instead, the only difference is the energy scale at which we observe the midpoint magnetization. The mean-field theory on a single ladder allows for highly accurate calculations of the magnetic phases and accurately captures the upper critical field $h_{c_2}$, in agreement with the exact solution we provide. Our results can be used to locate a large variety of spin ladder systems in the phase diagram and accurately determine the coupling parameters from magnetization data.

\section*{Acknowledgements} \label{sec:acknowledgements}
The authors thank Thierry Giamarchi, Miles Stoudenmire, Ryan Levy, Uli Schollwöck, John Sous, and David Reichman for helpful discussions. M.C.W. thanks Karl Pierce, Matthew Fishman and Géraud Krawezik for support with cluster calculations. 
M.C.W. acknowledges support from the German Academic Scholarship Foundation, the Swiss Study Foundation, and ETH Zurich. M.C.W. also acknowledges the hospitality of the Center for Computational Quantum Physics at the Flatiron Institute. The Flatiron Institute is a division of the Simons Foundation.

\section*{Contributions} \label{sec:contributions}
M.C.W. implemented the code, performed numerical simulations, and analyzed the data. M.C.W. and A.J.M. interpreted the data. M.C.W. wrote the manuscript with assistance from A.J.M. A.J.M. supervised the project.

\bibliography{ref}
\bibliographystyle{unsrt} 

\newpage
\onecolumngrid
\appendix
\section{Supplemental Material}
\counterwithin{figure}{section}
\subsection{Conversion of Units}
Our units can easily be converted into physical units, for example, by comparison with density functional theory (DFT) values for $t_i$ and $U$ via $J_i = \frac{4t_i^2}{U}$. In our calculations, we use units of $J_{\perp} = 1$ and $g\mu_B = 1$, where the Landé factor $g \approx 2$ depends on the experimental setup\cite{PhysRevB.41.1657}, and the Bohr magneton is $\mu_B \approx 5.79 \cdot 10^{-5}$ eV/T. We can then express the magnetic field in units of Tesla $[T]$:
\begin{equation}
    h_{\text{phys}}[\text{T}] = \frac{J_{\perp} [\text{eV}]}{g \cdot \mu_B [\text{eV/T}]} \cdot h_{\text{theory}}
\end{equation}
where $h_{\text{theory}}$ is the magnetic field in units used in this article. Another way is measuring the midpoint magnetization $h_m$, which is close to the dominating energy scale of $J_{\perp}$ for small intraladder coupling, see eq. \ref{eq:hm_small_gamma}, and close to $2\gamma$ for large intraladder coupling according to \ref{eq:hm_large_gamma}.

For a terphenyl compound described by Sous et al.\cite{sous2021}, we can use the following matrix elements obtained from DFT calculations: $t_{\perp} = 0.022\text{eV}$, $t_{\parallel} = 0.028\text{eV}$, and $t' = 0.013\text{eV}$ to calculate the Heisenberg couplings. In units of $t_{\perp} = 1$, we obtain the following ratios for the intraladder coupling $\gamma = (\frac{0.028}{0.022})^2 \approx 1.62$ and interladder coupling $J' \approx 0.35$. They also report estimates for the on-site Coulomb repulsion $U \approx 0.5 - 2\text{eV}$, which is much larger than all of the relevant hopping elements $t$, justifying the Heisenberg model as a valid description of this system at half-filling. Based on our calculations, we expect the ground state of the terphenyl system\cite{sous2021} to be an antiferromagnet. This state is expected to be stable up to $T_c \approx 4 - 16 \text{K}$ for $U \approx 2 - 0.5\text{eV}$\cite{PhysRevB.83.054407}.

\subsection{Upper Critical Field}\label{sec:appendix_hc2}
Similar to these previous works\cite{PhysRevB.55.3046, Chaboussant1998}, we calculate the exact value of the upper critical field $h_{c_2}$ for coupled spin ladders by determining the lowest energy of a single spin flip excitation. We define the fully polarized ground state
\begin{equation}
    \ket{\psi_\text{FP}} = \ket{\uparrow, \,..., \uparrow}
\end{equation}
as an eigenstate of our Hamiltonian in eq. \ref{eq:coupledspinladders}. The most general one-magnon excitation $M$ is
\begin{equation}\label{eq:one_magnon_state}
    \ket{\psi_{\text{FP}}^-} = \sum_{i=1}^{N_x}\sum_{j=1}^{N_y} \alpha_{ij} S^-_{ij} \ket{\psi_\text{FP}}  \equiv M \ket{\psi_{\text{FP}}} \,.
\end{equation}
The coefficient matrix $\alpha_{ij}$ is determined by demanding that $ \ket{\psi_{\text{FP}}^-}$ is an eigenstate of $H$ with energy $E^-$, meaning that
\begin{equation}
    H \ket{\psi_{\text{FP}}^-} = E^- \ket{\psi_{\text{FP}}^-}  \,.
\end{equation}
We know that
\begin{equation}
    H\left(M \ket{\psi_\text{FP}}  \right) = [H,M] \ket{\psi_\text{FP}} + M\left(H \ket{\psi_\text{FP}} \right)
\end{equation}
or, equivalently,
\begin{eqnarray}
    E^- \ket{\psi_\text{FP}^-}  &=& [H,M] \ket{\psi_\text{FP}} + E \ket{\psi_\text{FP}^-}
    \nonumber
    \\
    &&\Leftrightarrow (E - E^-) \ket{\psi_\text{FP}^-} 
    \nonumber\\
    &&= [M,H] \ket{\psi_\text{FP}} \,.
\end{eqnarray}
The commutator $[M,H]$ can be written as
\begin{equation}
    [M,H] = \sum_{i=1}^{N_x}\sum_{j=1}^{N_y} \alpha_{ij} [S^-_{ij},H]
\end{equation}
with terms
\begin{equation}
   [S^-_{ij},H_{\text{mag}} + H_{J_{\perp}} +  H_{J_{\parallel}, J'}]  \,.
\end{equation}
The magnetic contribution is a constant energy term
\begin{equation}
     [S^-_{ij},H_{\text{mag}}] = \sum_{k,l} h^z S^-_{kl}
\end{equation}
where $\sum_{k,l} = \sum_{k=1}^{N_x} \sum_{l=1}^{N_y}$ sums over all sites. The commutator along the legs leads to 
\begin{equation}
    [S^-_{ij},H_{J_{\perp}}] = \sum_{k,l} 2J_{\parallel} S^-_{k,l}S^z_{k+1,l}
\end{equation}
and along the $y$-direction we find
\begin{equation}
    [S^-_{ij},H_{J_{\parallel},J'}] = \sum_{k,l} 2J^y_l
    S^-_{k,l}S^z_{k,l+1}  \,.
\end{equation}

\subsubsection{Interladder Coupling on Every Rung}
In the following, we define the coupling in the second dimension for the BPCB system where every rung is coupled as
\begin{equation}\label{eq:Jy}
    J^y_j = J_{\perp}\delta_{\text{odd}(j)} + J' \delta_{\text{even}(j)} \,.
\end{equation}
This means that we have
\begin{equation}
    \begin{split}
        (E - E^-) \ket{\psi_\text{FP}^-} = 
        \sum_{i,j}
        \alpha_{ij} 
        \sum_{k,l}
        \big( 
         h^z S^-_{k,l}
         +
         2J_{\parallel} S^-_{k,l}S^z_{k+1,l}
         +
         2J^y_iS^-_{k,l}S^z_{k,l+1}
         \big) \ket{\psi_\text{FP}} \,.
    \end{split}
\end{equation}
By inserting eq. \ref{eq:one_magnon_state}, defining $\varepsilon \equiv E - E^-$, and applying $\bra{\psi_{\text{FP}}}$, we notice that this is equivalent to the following set of coupled real-space equations
\begin{equation}\label{eq:coupled_eq}
    \varepsilon \alpha_{ij} = h^z\alpha_{ij} 
    + 
    J_{\parallel} \left(-\alpha_{ij} + 
    \frac{\alpha_{i-1,j} + \alpha_{i+1,j}}{2} \right)
    + 
    \left(-\frac{J_{\perp} + J'}{2}\alpha_{ij} + 
    \frac{J^y_{j-1}\alpha_{i,j-1} + J^y_j\alpha_{i,j+1}}{2}\right) \,.
\end{equation}
Due to the translation invariance along $x$ and $y$ by one and two sites, respectively, we choose our coefficients on sublattices A,B
\begin{equation}
    a_{i,m} \equiv \alpha_{i,j=2m-1} \,,\,  b_{i,m} \equiv \alpha_{i,j=2m}
\end{equation}
where $a_{i,m}$ ($b_{i,m}$) represents odd (even) vertical bonds with $J_{\perp}$ ($J'$). We use two sets of plane waves 
\begin{equation}
    a_{i,m} = a e^{\mathrm{i}(k_x i + k_y m)} \,, \quad b_{i,m} = b e^{\mathrm{i}(k_x i + k_y m)}
\end{equation}
leading to two solutions for eq. \ref{eq:coupled_eq}. We define $\varepsilon(k_x,k_y) \equiv E - E^-(k_x,k_y)$ and get
the following momentum space dispersion
\begin{equation}\label{eq:dispersion}
    \varepsilon(k_x,k_y)= \varepsilon_{\text{mag}} + \varepsilon_{J_{\parallel}}(k_x) + \varepsilon_{J_{\perp},J'}(k_y)
\end{equation}
where
\begin{equation}
    \varepsilon_{\text{mag}} = h^z \,\quad \text{and} \quad\, \varepsilon_{J_{\parallel}}(k_x) = - J_{\parallel}(1-\cos{k_x})
\end{equation}
The term along the $y$-direction mixes momenta from different sublattices
\begin{equation}
    \varepsilon_{J_{\perp},J'}(k_y) = 
    \begin{cases}
        - \frac{J_{\perp} + J'}{2}a + \frac{J_{\perp}}{2}b + \frac{J'}{2}b e^{-\mathrm{i}k_y} & \,\,\text{on A}\\
        - \frac{J_{\perp} + J'}{2}b + \frac{J_{\perp}}{2}a + \frac{J'}{2}a e^{\mathrm{i}k_y} & \,\,\text{on B}
    \end{cases}
\end{equation}
leading to the following eigenproblem for eq. \ref{eq:dispersion}
\begin{equation}\label{eq:eigenproblem}
    \varepsilon(k_x,k_y)
    \begin{pmatrix}
        a \\ b
    \end{pmatrix} = 
    \Big( 
    \varepsilon_{\text{mag}} + \varepsilon_{J_{\parallel}}(k_x)
    \Big) 
    \begin{pmatrix}
        a \\ b
    \end{pmatrix}
    +
    \begin{pmatrix}
        - \frac{J_{\perp} + J'}{2} & \frac{J_{\perp} + J'e^{-\mathrm{i}k_y}}{2} \\
        \frac{J_{\perp} + J'e^{\mathrm{i}k_y}}{2}  & - \frac{J_{\perp} + J'}{2}
    \end{pmatrix}
    \begin{pmatrix}
        a \\ b
    \end{pmatrix}
\end{equation}
which is equivalent to
\begin{equation}\label{eq:dispersion_pm}
    \varepsilon(k_x,k_y)_{\pm} 
    = 
    h^z - J_{\parallel}(1-\cos{k_x})
    - \frac{J_{\perp} + J'}{2} \pm \frac{1}{2}\sqrt{J_{\perp}^2 + J'^2 + 2J_{\perp} J' \cos{k_y}} \,.
\end{equation}
The critical field $h_{c_2}$ is specified by the condition that the lowest one-magnon mode, which is contained in $\varepsilon(k_x,k_y)_{-}$, has zero energy 
\begin{equation}\label{eq:min_magnon_energy}
    \text{min}_{k_x, k_y} \varepsilon(k_x,k_y)_{-} = 0 \,.
\end{equation}
This is achieved when $1-\cos{k_x}$ is maximal at $k_x = \pi$, and $\sqrt{J_{\perp}^2 + J'^2 + 2J_{\perp} J' \cos{k_y}}$ is maximal at $k_y = 0$
\begin{equation}
    0 \overset{!}{=} \varepsilon(k_x = \pi,k_y = 0)_{-} = h^z - J_{\parallel}(1-\cos{\pi}) - 
    \frac{J_{\perp} + J'}{2} - \frac{\abs{J_{\perp} + J'}}{2}
\end{equation}
leading to
\begin{equation}\label{eq:hc2_coupled_ladders}
    h_{c_2} = 2J_{\parallel} + J_{\perp} + J'
\end{equation}
\subsubsection{Interladder Coupling on Every Second Rung}
For a system such as terphenyl, where only every second rung is coupled, we have to modify eq. \ref{eq:Jy} as follows:
\begin{equation}\label{eq:Jy_terphenyl}
    J^y_{i,j} = J_{\perp}\delta_{\text{odd}(j)} + J' \delta_{\text{even}(i)}\delta_{\text{even}(j)} \,.
\end{equation}
This means that our set of coupled equations becomes
\begin{equation}\label{eq:coupled_eq_2}
    \varepsilon \alpha_{ij} = h^z\alpha_{ij} 
    + 
    J_{\parallel} \left(-\alpha_{ij} + 
    \frac{\alpha_{i-1,j} + \alpha_{i+1,j}}{2} \right)
    + 
    \left(-\frac{J^y_{i,j} + J^y_{i,j-1}}{2}\alpha_{ij} + 
    \frac{J^y_{i,j-1}\alpha_{i,j-1} + J^y_{i,j}\alpha_{i,j+1}}{2}\right)
\end{equation}
and our unit cell now contains four sites. We use the following plane wave basis for even and odd sites $i$ along $x$ where we have alternating ($J_{\perp},J'$)- and ($J_{\perp},0$)-bonds, respectively:
\begin{equation}
\begin{split}
    a_{\text{even},n,m} \equiv \alpha_{i=2n,j=2m-1} \,,\,&  b_{\text{even},n,m} \equiv \alpha_{i=2n,j=2m} \\   
    a_{\text{odd},n,m} \equiv \alpha_{i=2n-1,j=2m-1} \,,\,&  b_{\text{odd},n,m} \equiv \alpha_{i=2n-1,j=2m} \\
\end{split}
\end{equation}
The dispersion resulting from eq. \ref{eq:coupled_eq_2} along $x$ remains unchanged. Similarly, we have the same result along even $y$-bonds, leading to the same eigenproblem as in eq. \ref{eq:eigenproblem}. However, along odd $y$-bonds, we have:
\begin{equation}\label{eq:eigenproblem_odd_bonds}
    \varepsilon(k_x,k_y)_{\text{odd}}
    \begin{pmatrix}
        a_{\text{odd}} \\ b_{\text{odd}}
    \end{pmatrix} = 
    \Big( 
    \varepsilon_{\text{mag}} + \varepsilon_{J_{\parallel}}(k_x)
    \Big) 
    \begin{pmatrix}
        a_{\text{odd}} \\ b_{\text{odd}}
    \end{pmatrix}
    +
    \begin{pmatrix}
        - \frac{J_{\perp}}{2} & \frac{J_{\perp}}{2} \\
        \frac{J_{\perp}}{2}  & - \frac{J_{\perp}}{2}
    \end{pmatrix}
    \begin{pmatrix}
        a_{\text{odd}} \\ b_{\text{odd}}
    \end{pmatrix}
\end{equation}
with eigenvalues $\{-J_{\perp},0\}$ for the second matrix. In total, we have four solutions for the dispersion, two for even bonds:
\begin{equation}
    \varepsilon(k_x,k_y)_{\pm,\text{even}} = \varepsilon(k_x,k_y)_{\pm} \,,
\end{equation}
which is the same result we obtained in eq. \ref{eq:dispersion_pm}, and two for odd bonds:
\begin{equation}
    \varepsilon(k_x)_{\pm,\text{odd}} 
    = 
    h^z - J_{\parallel}(1-\cos{k_x})
    - \frac{J_{\perp} \pm J_{\perp}}{2} \,,
\end{equation}
which are independent of $k_y$. 
We again search for the global minimum of the one-magnon energy and set it to zero, which is at $k_x = \pi$. For $k_y$, we have to compare the minimum energies of $\varepsilon(\pi,k_y)_{\pm,\text{even}}$ and  $\varepsilon(\pi)_{\pm,\text{odd}}$, which are
\begin{equation}
    \varepsilon(\pi,k_y = \pi)_{-,\text{even}} = h^z - 2J_{\parallel} - J_{\perp} - J' 
\end{equation}
and 
\begin{equation}
    \varepsilon(\pi)_{+,\text{odd}} = h^z - 2J_{\parallel} - J_{\perp}
\end{equation}
respectively. Hence, the minimum one-magnon energy is contained in $\varepsilon(k_x = \pi,k_y = 0)_{-,\text{even}}$ on even rung sites $i$ where we have alternating ($J_{\perp}, J'$)-couplings. Removing interladder couplings $J'$ does not change the minimum in eq. \ref{eq:min_magnon_energy}, and therefore, the upper critical field remains the same. 

For $J' = 0$, we recover the exact solution $h_{c_2} = J_{\perp} + 2J_{\parallel}$ for a single spin ladder\cite{PhysRevB.55.3046, Chaboussant1998}. Our result in eq. \ref{eq:hc2_coupled_ladders} is general for any number of rungs that are not coupled between individual spin ladders, as long as we have at least one single chain of alternating $(J_{\perp}, J')$-couplings in the second dimension.  

\subsection{DMRG Calculations}
We performed DMRG calculations with a maximum bond dimension of $m=500$, although most MPS can be represented with $m<10$ in the spin-gapped and fully polarized phases. Close to the phase transition, the ground states show strong entanglement, and we obtain MPS with the maximum bond dimension and truncation errors of $\varepsilon \approx 10^{-7}$. Although all of our DMRG calculations have been performed with periodic boundary conditions, we find good convergence with 40 DMRG sweeps starting from a random initial state and keeping states up to an SVD truncation error cutoff of $\varepsilon = 10^{-10}$. 

\subsection{Finite Size Scaling}\label{sec:appendix_finite_size}
\begin{figure}[htbp!]
 \centering
 \begin{adjustbox}{center}
   \includegraphics[width=1\columnwidth]{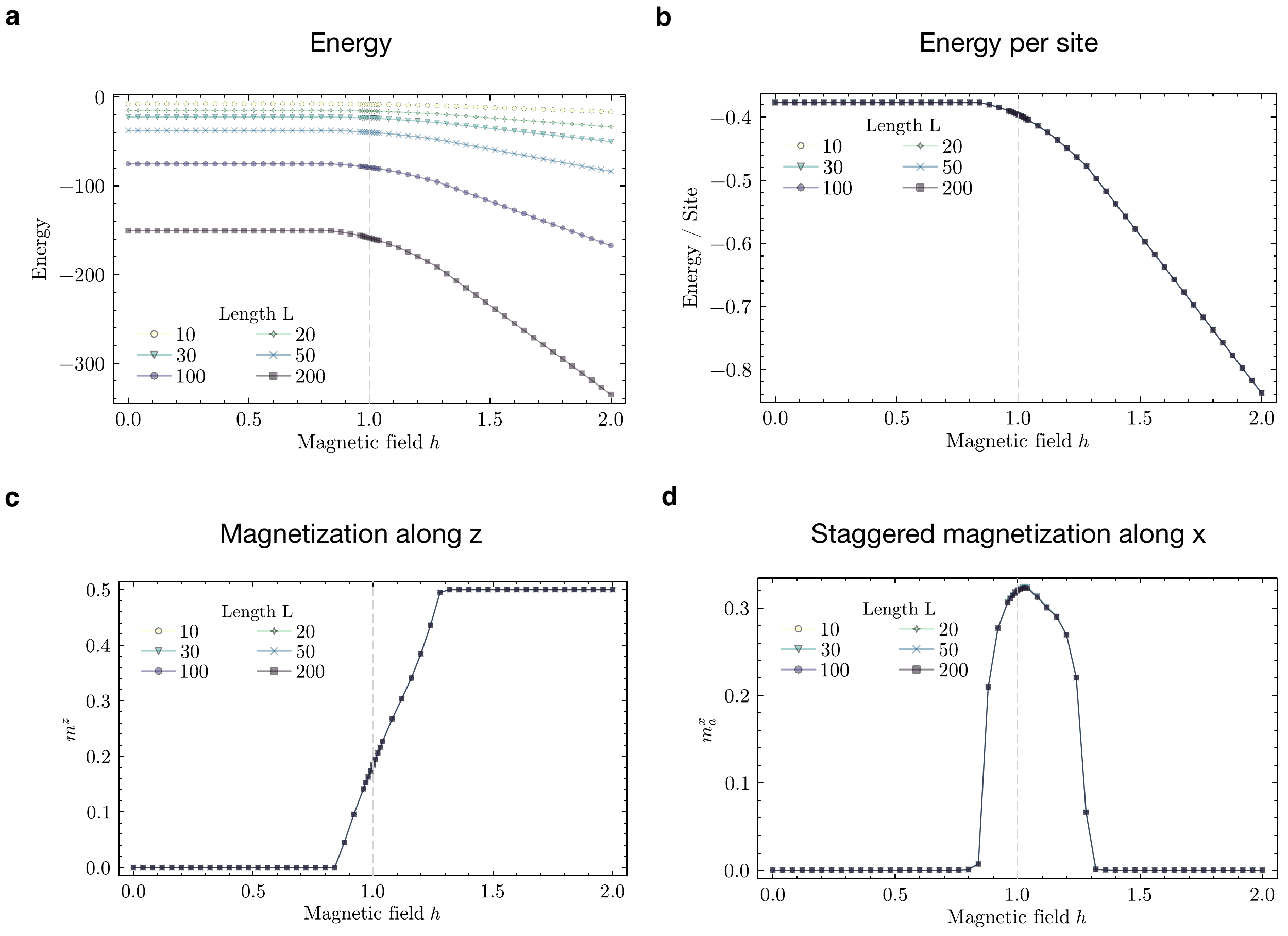}
 \end{adjustbox}
 \caption{Finite-size study of mean-field spin ladder coupled on every second rung with small intraladder coupling $\gamma = 0.1$ and interladder coupling $J' = 0.1$ for different magnetic fields $h$ and different lengths between $L = 10$ and $L=200$. (a) Ground state energy. (b) Ground state energy normalized by the number of sites. (c) Magnetization along $z$ ($m^z$). (d) Staggered magnetization along $x$ ($m^x_a$).}
 \label{fig:fig_finite_size_01}
\end{figure}

\begin{figure}[htbp!]
 \centering
 \begin{adjustbox}{center}
   \includegraphics[width=1\columnwidth]{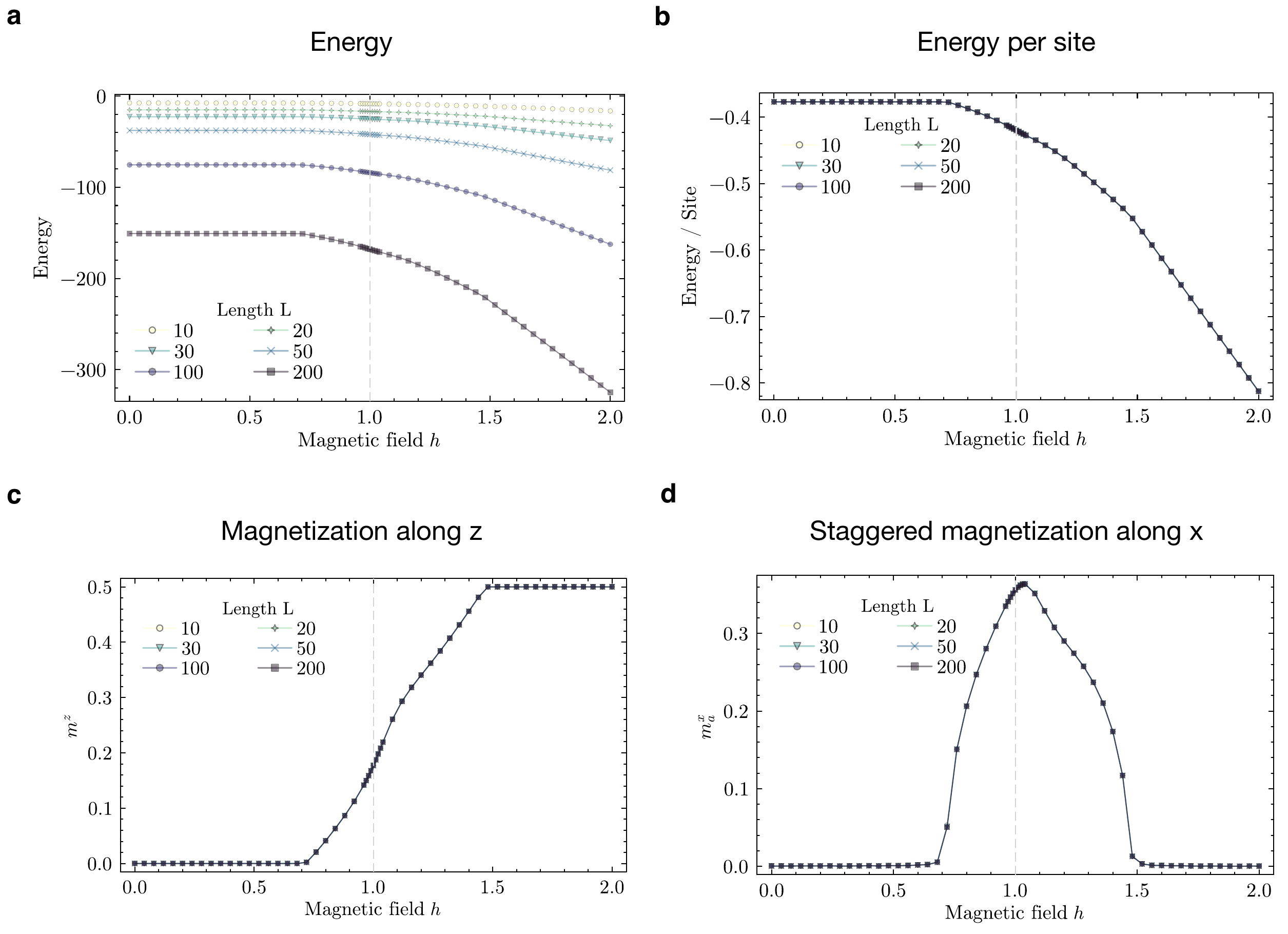}
 \end{adjustbox}
 \caption{Finite-size study of mean-field spin ladder coupled on every second rung with intraladder coupling $\gamma = 0.1$ and interladder coupling $J' = 0.3$ for different magnetic fields $h$ and different lengths between $L = 10$ and $L=200$. (a) Ground state energy. (b) Ground state energy normalized by the number of sites. (c) Magnetization along $z$ ($m^z$). (d) Staggered magnetization along $x$ ($m^x_a$).}
 \label{fig:fig_finite_size_02}
\end{figure}

\begin{figure}[htbp!]
 \centering
 \begin{adjustbox}{center}
   \includegraphics[width=1\columnwidth]{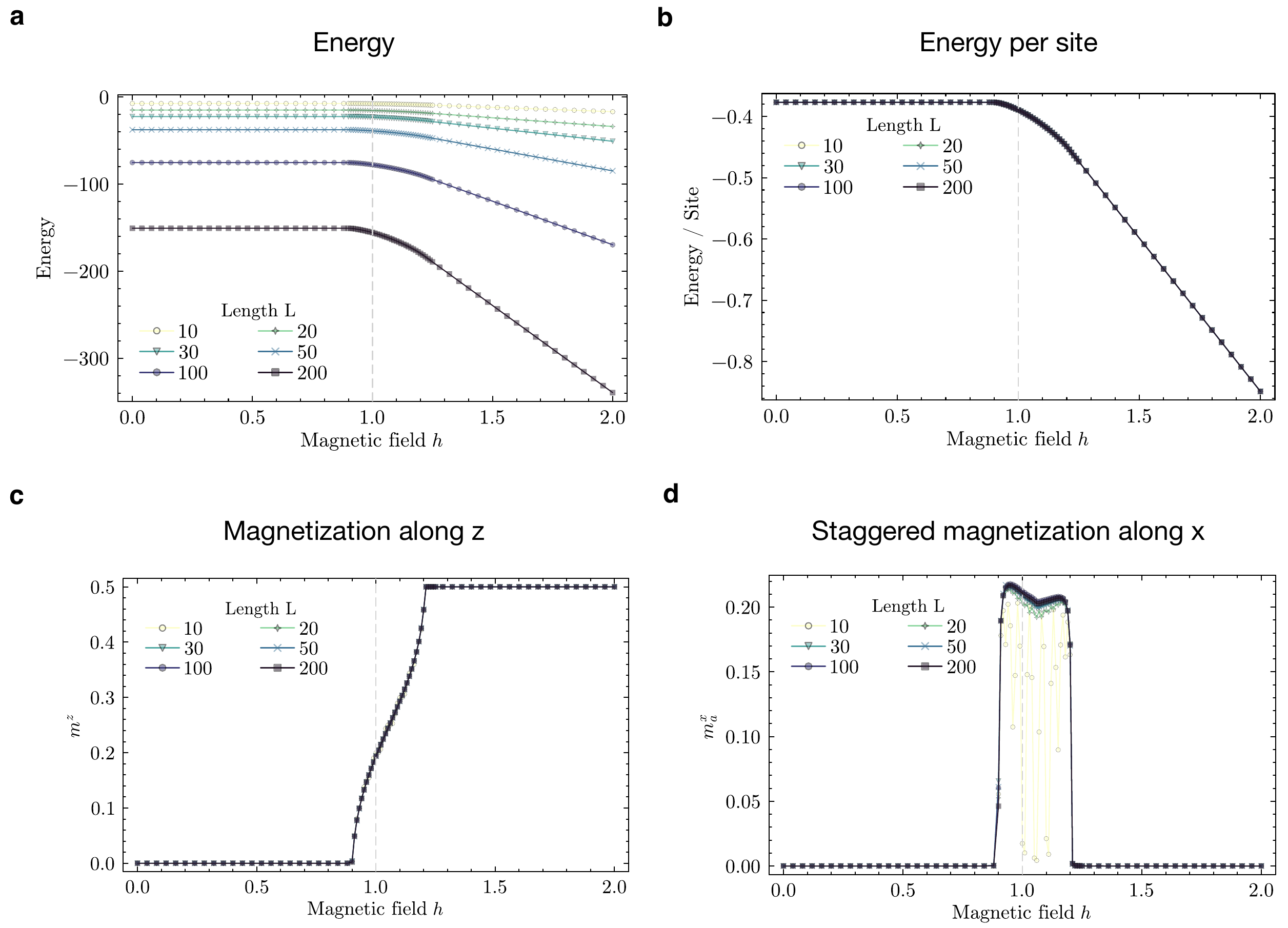}
 \end{adjustbox}
 \caption{Finite-size study of mean-field spin ladder coupled on every second rung with intraladder coupling $\gamma = 0.1$ and interladder coupling $J' = 0.01$ for different magnetic fields $h$ and different lengths between $L = 10$ and $L=200$. (a) Ground state energy. (b) Ground state energy normalized by the number of sites. (c) Magnetization along $z$ ($m^z$). (d) Staggered magnetization along $x$ ($m^x_a$).}
 \label{fig:fig_finite_size_03}
\end{figure}

In order to estimate the influence of limited system size, we performed a finite-size calculation for spin ladders coupled in mean-field on every second rung of length between $L=10$ and $L=200$ with different interladder couplings $J' = 0.1$ and $J' = 0.3$, see figs. \ref{fig:fig_finite_size_01} and \ref{fig:fig_finite_size_02}, respectively. All calculations were performed with an intraladder coupling of $\gamma = 0.1$ and with periodic boundary conditions. Remarkably, even a system of length $L=10$ captures the qualitative physics without significant deviations, and a system of length $L=30$ is nearly indistinguishable from $L=200$. In fig. \ref{fig:fig_finite_size_03}, we show that our mean-field theory predicts a staggered order in the gapless region even for $J'=0.01$, which is not due to finite size effects. Again, we observe that a system of length $L=30$ is a good description of the infinite size limit.

\begin{figure}[htbp!]
 \centering
 \begin{adjustbox}{center}
   \includegraphics[width=1\columnwidth]{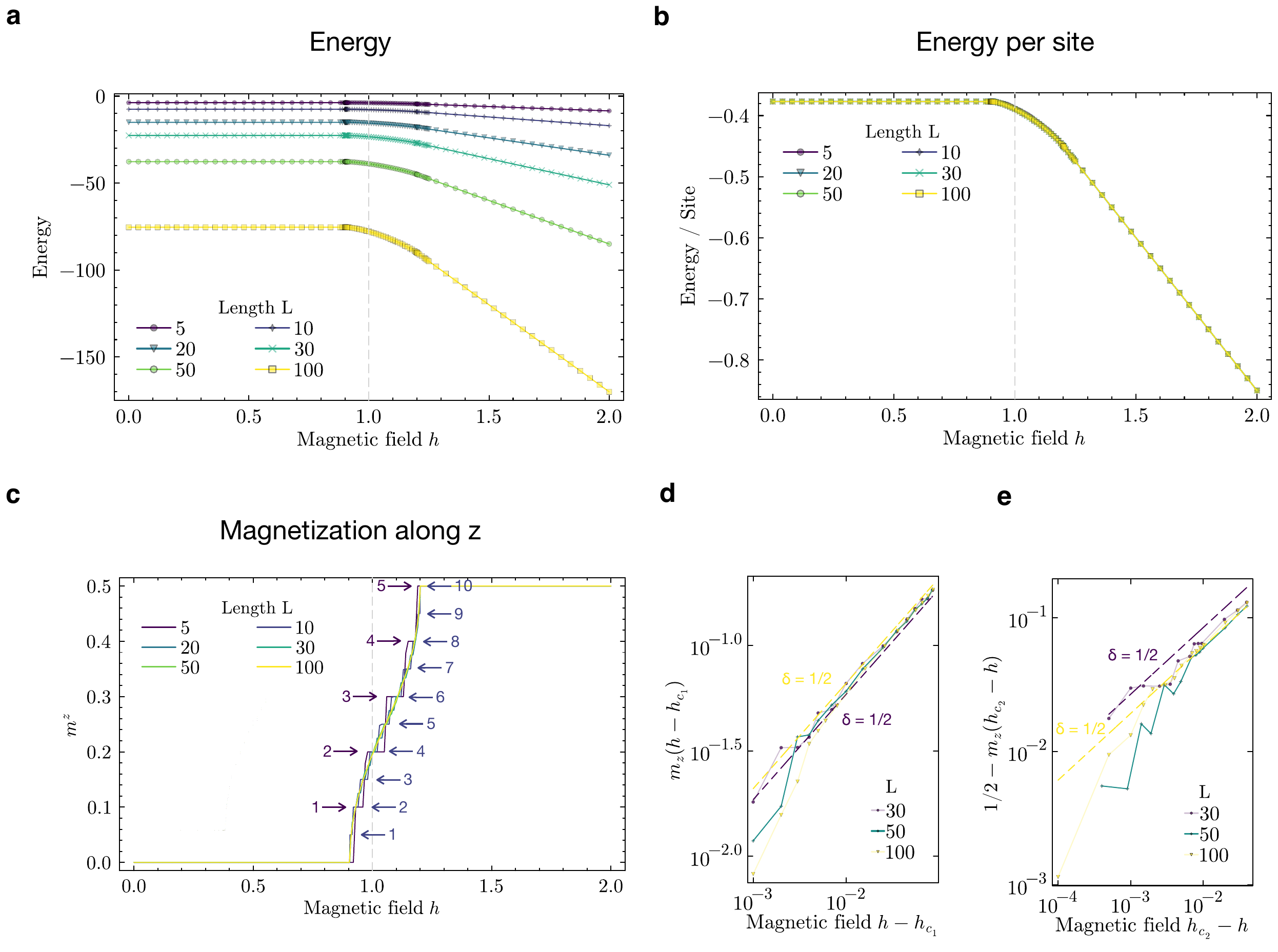}
 \end{adjustbox}
 \caption{Finite-size study of single spin ladder with intraladder coupling $\gamma = 0.1$ for different magnetic fields $h$ and different lengths between $L = 10$ and $L=100$. (a) Ground state energy. (b) Ground state energy normalized by the number of sites. (c) Magnetization along $z$ ($m^z$). (d) Critical scaling close to $h_{c_1}$. Dashed lines indicate square root scaling with $\delta = 1/2$. (e) Critical scaling close to $h_{c_2}$.}
 \label{fig:fig_finite_size_04}
\end{figure}

In fig. \ref{fig:fig_finite_size_04}, we present a finite size study for a single spin ladder with intraladder coupling $\gamma = 0.1$ without mean-field couplings. The ground state energy is captured well even in small systems, see fig. \ref{fig:fig_finite_size_04}(b). In fig. \ref{fig:fig_finite_size_04}(c), we present the ground state magnetization. We observe a finite size effect in the magnetization curve, which shows $L$ steps in the gapless phase for a system of length $L$. We observe that, with increasing magnetic field $h$, individual rung singlet pairs are excited to triplet pairs.  We can see that the number of magnetization plateaus equals the number of rungs. The five (ten) different plateaus for $L=5$ ($L=10$) are highlighted with arrows on the left (right) side of the magnetization curve. We can observe that every second plateau for $L=10$ overlaps well with the plateau for $L=5$. This leads to a minimum system size of $L_{min} \gtrsim \frac{3\gamma}{\Delta h}$ to avoid noticeable finite size effects in the shape of the magnetization curve for a spin ladder, based on the size of the gapless region $3 \gamma$ and the magnetization grid size $\Delta h$. Note that the gapless region $h_{c_2} - h_{c_1}$ is smaller than $3\gamma$ for larger intraladder coupling. The critical scaling of a single spin ladder is visualized in fig. \ref{fig:fig_finite_size_04}(d) and (e) for the lower and upper critical fields, respectively. One can clearly observe the square root scaling with $\delta = 1/2$, see also eq. \ref{eq:critscaling_sqrt}. The data for $L=100$ is used in fig. \ref{fig:fig3}(c) as the curve for $J' = 0$.

The magnetization plateaus are also the source of finite size effects in fig. \ref{fig:fig4}(b) and (d). In these calculations, we use spin ladders with large intraladder couplings $\gamma = 1.5$ and $\gamma = 3$, respectively. While both curves for the staggered magnetization are smooth for $J' > 0.2$, we notice "wobbles" for $J'=0.1$ and $\gamma = 3$, which become more pronounced as peaks for $J'=0.01$. It turns out that, due to limited system size, the numerical mean-field approximation only converges to AFM order when the magnetic field $h$ is approximately a multiple of the minimum field necessary to flip a rung pair $\frac{\Delta}{L}$, where $\Delta = h_{c_2} - h_{c_1}$ is the width of the gapless region and $L$ is the number of rungs. For limited system sizes, there is only a small range of magnetic fields that is large enough to enable the triplet mobility necessary to induce staggered order but not so large as to flip rung pairs and move up to the next magnetization plateau. The infinite size limit will follow the envelope of $m^x_a$, see also fig. \ref{fig:fig_finite_size_03}, where we observe the same finite-size effect in a small system of length $L=10$ for $J' = 0.01$. In the case of very large intraladder coupling $\gamma$ and/or very small interladder coupling $J'$, one may use larger systems to perform accurate simulations of magnetization curves. We find that the occurrence of finite size effects can be easily estimated.

\subsection{Effect of Intraladder Coupling $\gamma$}
\begin{figure}[htbp!]
 \centering
 \begin{adjustbox}{center}
   \includegraphics[width=1\columnwidth]{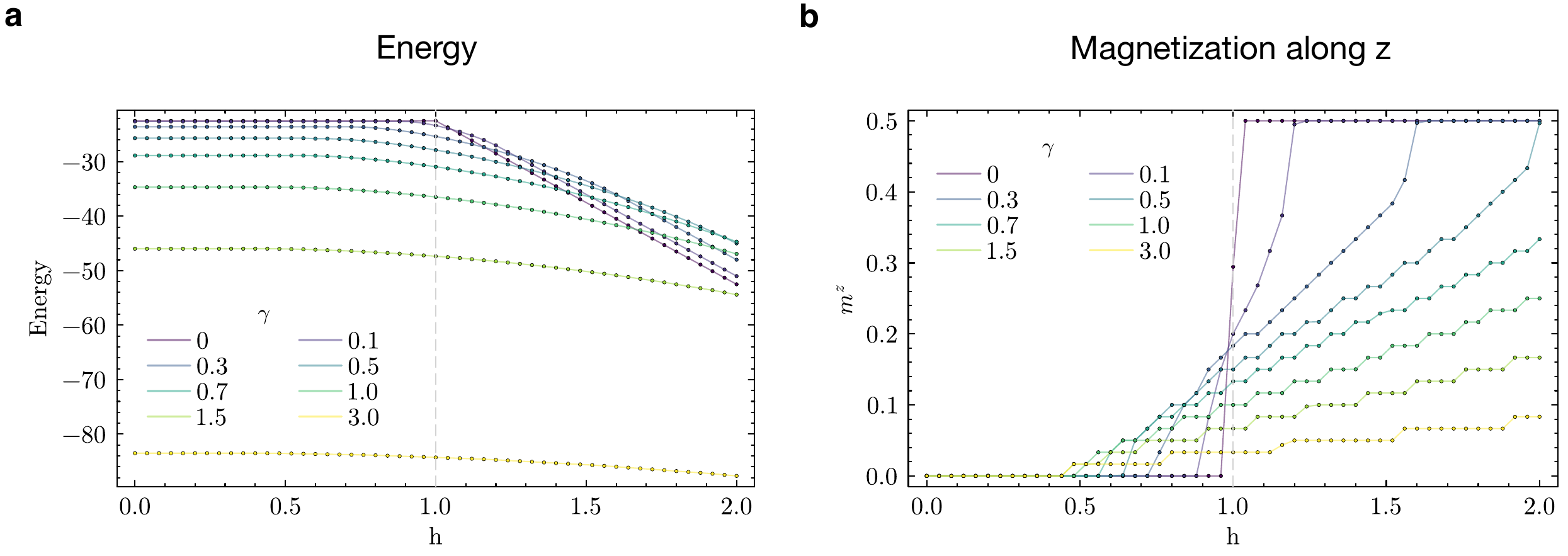}
 \end{adjustbox}
 \caption{Single spin ladder of length $L=30$ with different intraladder couplings $\gamma$. (a) Ground state energy. (b) Magnetization along $z$. Note that the magnetization plateaus are due to finite size effects.}
 \label{fig:figures_SI_single_spin_ladder}
\end{figure}

For a single spin ladder, there is always a spin gap that depends on the intraladder coupling $\gamma$, see fig. \ref{fig:figures_SI_single_spin_ladder}(a) and (b), where we present the ground state energy and magnetization along $z$ for a single spin ladder of length $L=30$. The magnetization plateaus are due to finite size effects, see fig. \ref{fig:fig_finite_size_04}, which do not change the spin gap. In the decoupled rung limit of $\gamma = 0$, the spin gap is equal to $J_{\perp} = 1$. For small gamma up to $\gamma \approx 0.3$, we can use the lower critical field $h_{c_1} = J_{\perp} - J_{\parallel}$ to estimate the spin gap. At $\gamma = 1$, the spin gap is approximately $\gamma / 2 = 0.5$. For larger $\gamma$, the gap decreases and the system becomes gapless in the limit of $J_{\perp} = 0$\cite{PhysRevB.47.3196}. Note that we use units of $J_{\perp} = 1$, in which the gap is $\Delta \approx J_{\perp}/2$ for larger $\gamma$\cite{PhysRevB.69.092408}.

\subsection{Convergence of Numerical Mean-Field Theory}
\begin{figure}[htbp!]
 \centering
 \begin{adjustbox}{center}
   \includegraphics[width=1\columnwidth]{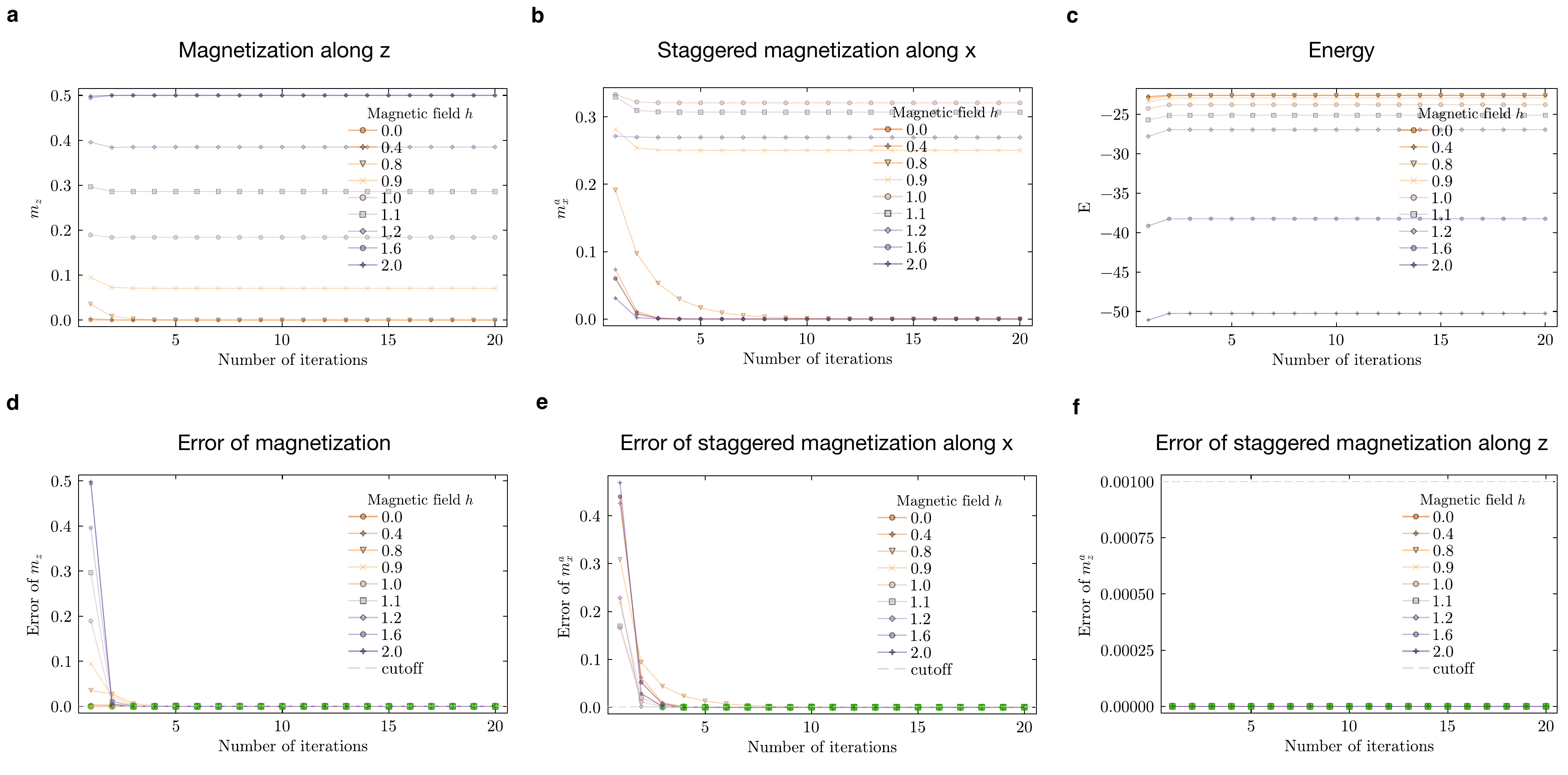}
 \end{adjustbox}
 \caption{Convergence of observables with iterations of numerical mean-field theory for a spin ladder coupled on every second rung with $\gamma = 0.1$ and $J' = 0.1$. (a) Magnetization along $z$. (b) Staggered magnetization along $x$. (c) Ground state energy. (d) - (f) Convergence of magnetization, staggered magnetization along $x$, and along $z$, respectively.}
 \label{fig:fig_convergence_1}
\end{figure}

\begin{figure}[htbp!]
 \centering
 \begin{adjustbox}{center}
   \includegraphics[width=1\columnwidth]{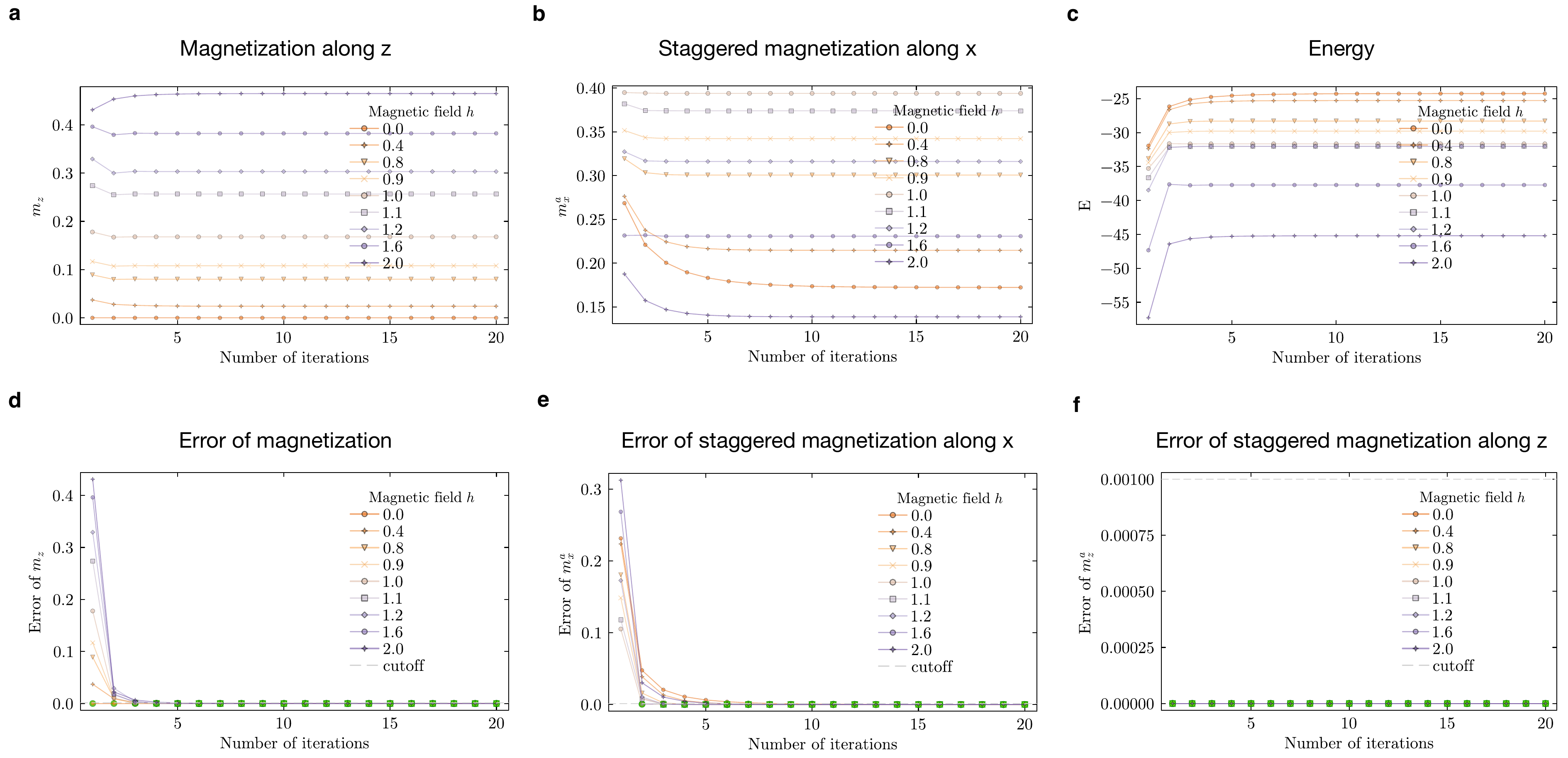}
 \end{adjustbox}
 \caption{Convergence of observables with iterations of numerical mean-field theory for a spin ladder coupled on every second rung with $\gamma = 0.1$ and larger $J' = 1.0$. (a) Magnetization along $z$. (b) Staggered magnetization along $x$. (c) Ground state energy. (d) - (f) Convergence of magnetization, staggered magnetization along $x$, and along $z$, respectively.}
 \label{fig:fig_convergence_2}
\end{figure}

Our numerical mean-field theory usually converges within a few iterations, although more iterations ($>10$) are needed close to the phase transitions. In figs. \ref{fig:fig_convergence_1} and \ref{fig:fig_convergence_2}, we present the observables of interest as a function of the iterations for different magnetic fields for a spin ladder of $L=30$ coupled on every second rung. We observe that the magnetization (a) and staggered magnetization (b) converge within a few iterations (c) - (d), as indicated by the green symbols. Close to the transition from staggered order to unpolarized, we need more iterations to reach convergence, see for example fig. \ref{fig:fig_convergence_2}(b) for $h=0$. For all calculations, we verified that there is no alternating order along $z$. We find that $m^z_a$ is very small, often of the order of $10^{-8}$ or less.

\subsection{Phase Diagrams}
\begin{figure}[htbp!]
 \centering
 \begin{adjustbox}{center}
   \includegraphics[width=1\columnwidth]{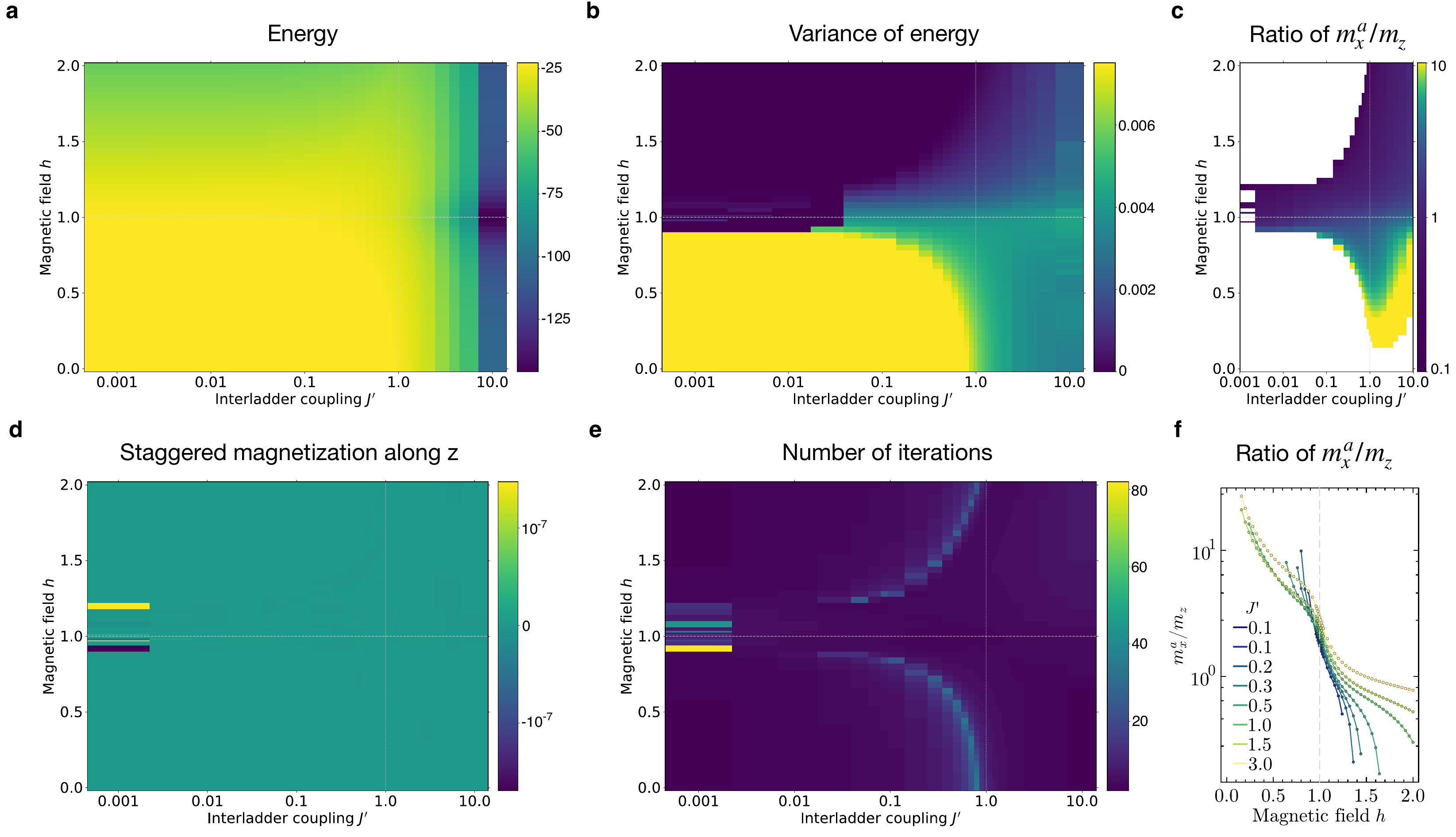}
 \end{adjustbox}
 \caption{Ground state phase diagram of a single spin ladder with small intraladder coupling $\gamma = 0.1$ treated with mean-fields on every other rung for different magnetic fields $h$ (vertical axis) and different interladder couplings $J'$ (horizontal axis). (a) Ground state energy. (b) Variance of ground state energy. (c) Ratio of staggered magnetization versus magnetization as a heatmap shows clear transition at $h = 1$. (d) Staggered magnetization along $z$. (e) Number of iterations until convergence is reached. (f) Ratio of $m^x_a / m^z$ as a function of magnetic field $h$ for different interladder couplings $J'$.}
 \label{fig:fig_phase_diagram_SI}
\end{figure}

In fig. \ref{fig:fig_phase_diagram_SI}(a) and (b), we present the ground state energy for the calculations shown in fig. \ref{fig:fig2}, together with the variance of the ground state energy $E_{var}$. This quantity indicates how well an MPS approximates the ground state of a given Hamiltonian. We calculate 
\begin{equation}
    E_{var} = \langle \hat{H}^2 \rangle - \langle \hat{H} \rangle^2
\end{equation}where $\langle ... \rangle$ denotes the ground state expectation value, computed with the ground state MPS, and $\hat{H}$ the respective Hamiltonian in Matrix Product Operator (MPO) form. Figure \ref{fig:fig_phase_diagram_SI}(d) proves that the staggered order along $z$ is of the order of $10^{-7}$ or lower and thus negligible. The number of iterations to reach convergence is shown in fig. \ref{fig:fig_phase_diagram_SI}(e). This number is usually below 10 for most parameters; however, close to the phase transition, it increases. Additionally, we present the ratio of staggered order ($m^x_a$) to magnetization along $z$ ($m^z$) in fig. \ref{fig:fig_phase_diagram_SI}(c), which can be used to infer whether a system has interladder couplings on every rung or not. In fig. \ref{fig:fig_phase_diagram_SI}(f), we observe a sharp decline of $m^x_a/m^z(h)$ close to $h \approx 1$, which is the energy scale of a single rung $J_{\perp} = 1$. This feature can be used to infer the presence of rungs without interladder couplings. 

\begin{figure}[htbp!]
 \centering
 \begin{adjustbox}{center}
   \includegraphics[width=1\columnwidth]{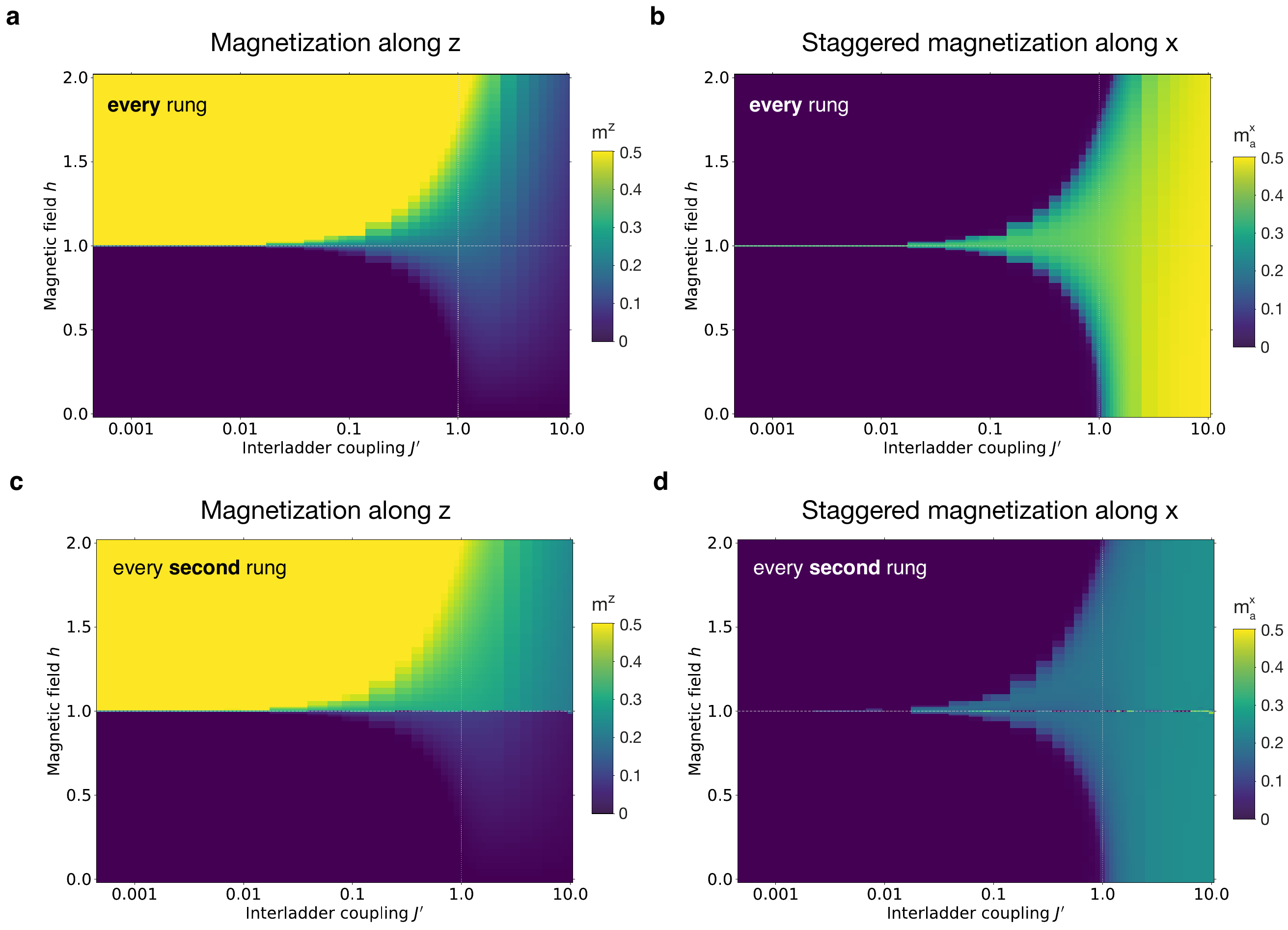}
 \end{adjustbox}
  \caption{Ground state phase diagram of a spin ladder in decoupled rung limit ($\gamma = 0$) treated with mean-fields on every rung (top row) and every second rung (bottom) for different magnetic fields $h$ (vertical axis) and different interladder couplings $J'$ (horizontal axis). (a) and (c): Magnetization along $z$. (b) and (d): Staggered magnetization along $x$.}
 \label{fig:fig_phase_diagram_gamma_0}
\end{figure}

In fig. \ref{fig:fig_phase_diagram_gamma_0}, we present the phase diagram for a spin ladder where either every rung is coupled in mean-field, see fig. \ref{fig:fig_phase_diagram_gamma_0}(a) and (b), or every second rung, fig. \ref{fig:fig_phase_diagram_gamma_0}(c) and (d), in the single rung limit of $\gamma = 0$. Note that this phase diagram is incorrect at $J' = 1$ where the actual system corresponds to a Heisenberg spin chain. However, we can use this model to estimate the phase boundaries in the limit of very weak intraladder coupling. These phase diagrams can be computed efficiently since we only need two (four) sites for a system where every (every other) rung is coupled.
We observe that the phase boundaries in this limit do not depend on the number of coupled rungs; however, the absolute magnitude of staggered order does. Since a single rung has a gap of $J_{\perp}$, we know that a system of decoupled rungs, where only every second is coupled, will show no magnetization along $z$ for the uncoupled rungs for $h < h_{c_1} = J_{\perp}$. We also know that these ground state excitations do not change qualitatively if we add a small non-zero coupling $J_{\parallel} \neq 0$ along the legs\cite{PhysRevB.83.054407}.

\subsection{Magnetization Curves}
\begin{figure}[htbp!]
 \centering
 \begin{adjustbox}{center}
   \includegraphics[width=1\columnwidth]{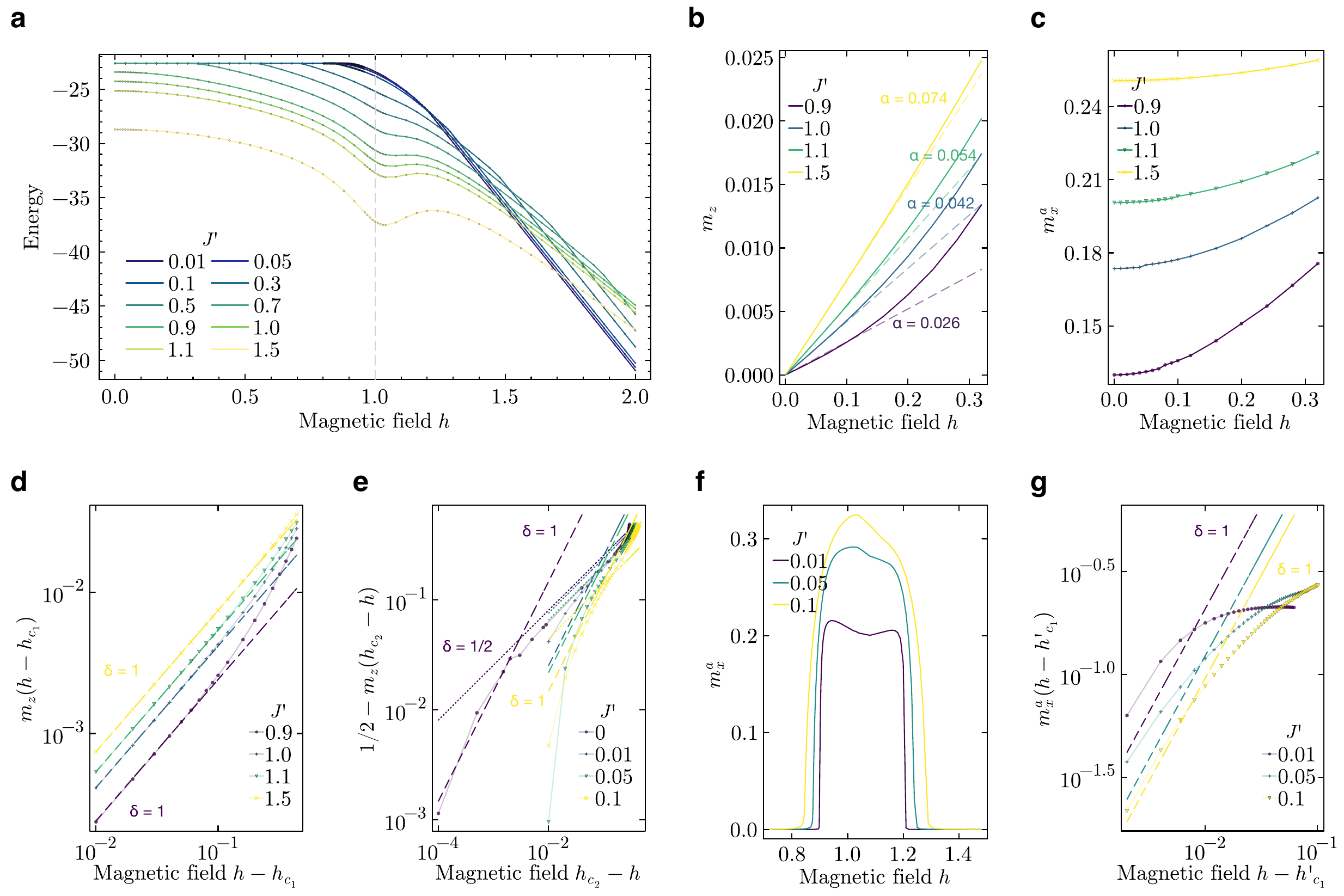}
 \end{adjustbox}
 \caption{Calculations for a spin ladder with small intraladder coupling $\gamma = 0.1$ and mean-field coupling on every second rung. (a) Ground state energy for different interladder couplings $J'$. (b) Magnetization along $z$ for larger $J'$ shows linear scaling close to $h\rightarrow 0 = h_{c_1}$ with different slopes $\alpha$. (c) Staggered magnetization along $x$ becomes constant close to $h\rightarrow 0$. (d) Critical scaling for large $J'$ close to zero field shows linear curves with different slopes $\alpha$. The critical field $\tilde{h}_{c_1}$ at which the slope changes increases with $J'$. (e) Critical scaling for small $J'$ close to the upper critical field $h_{c_2}$. $J' = 0$ denoted a single spin ladder without mean-field coupling. Note that data below $10^{-2}$ is unreliable due to finite-size effects. (f) Staggered magnetization along $x$ for small $J'$ shows increasing critical fields $h'_{c_1}$ and $h'_{c_2}$ with $J'$. (g) Critical scaling of $m^x_a$ close to $h'_{c_1}$ reveals linear scaling ($\delta = 1$) for $J'=0.1$.}
 \label{fig:fig3_SI_1}
\end{figure}

In fig. \ref{fig:fig3_SI_1}, we present additional data for the calculations from fig. \ref{fig:fig3}, including the ground state energy, see panel (a). Note that this quantity is missing the constant energy shift $\langle S^\alpha_{i,j} \rangle  \langle S^\alpha_{i',j'} \rangle$ from the mean-field decoupling for $S^\alpha_{i,j} S^\alpha_{i',j'} \rightarrow S^\alpha_{i,j} \langle S^\alpha_{i',j'} \rangle + \langle S^\alpha_{i,j} \rangle S^\alpha_{i',j'} - \langle S^\alpha_{i,j} \rangle  \langle S^\alpha_{i',j'} \rangle$, which is why $E(h)$ is not monotonically decreasing with magnetic field $h$. 
Figure \ref{fig:fig3_SI_1}(b) illustrates the linear scaling of $m^z$ close to $h\rightarrow 0$, where the slope can be used to distinguish different $J'$, see fig. \ref{fig:fig3_SI_1}(d). Close to zero field, the staggered order saturates to a constant value, which increases with $J'$, as shown in fig. \ref{fig:fig3_SI_1}(c). In fig. \ref{fig:fig3_SI_1}(e), we show that the square root ($\delta = 1/2$) and linear ($\delta = 1$) scaling of a single spin ladder with $J'=0$ and coupled spin ladders with $J' = 0.01, 0.05$ and $0.1$, respectively, also hold for the upper critical field $h_{c_2}$. The staggered order, presented in fig. \ref{fig:fig3_SI_1}(f) for small interladder couplings $J'$, also shows a linear scaling with $\delta = 1$ close to the lower critical field $h'_{c_1}$, similar to fig. \ref{fig:fig3}(d).

\begin{figure}[htbp!]
 \centering
 \begin{adjustbox}{center}
   \includegraphics[width=0.75\columnwidth]{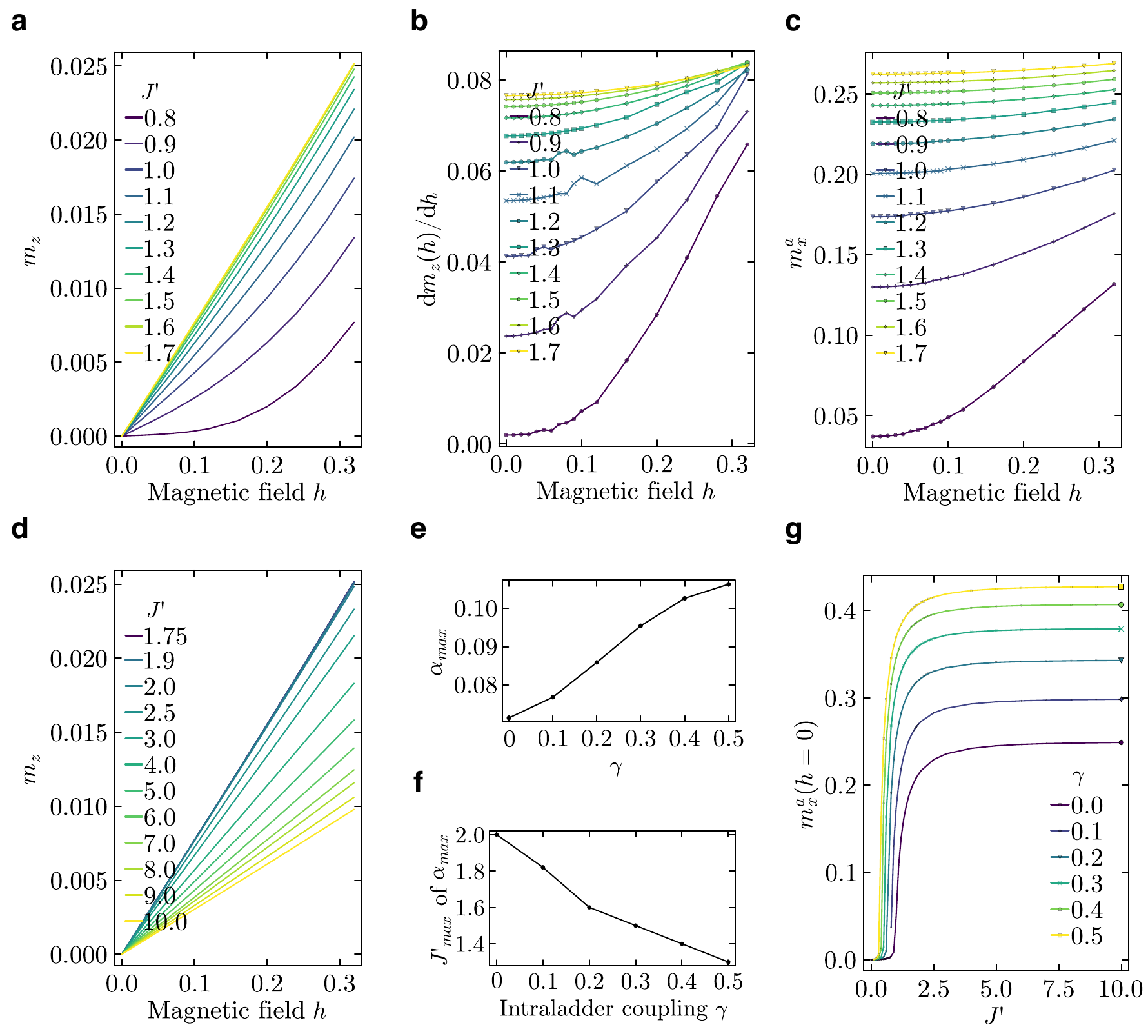}
 \end{adjustbox}
 \caption{Additional calculations for a spin ladder with small intraladder coupling $\gamma = 0.1$ and mean-field coupling on every second rung. (a) Magnetization along $z$ ($m^z$) shows linear behavior close to zero field for larger interladder couplings $J'$. The slope increases with $J'$ up to $J'_{max} \approx 1.7$. (b) The slope $\alpha$ of $m^z$ increases with $J'$, similar to $\tilde{h}_{c_1}$. (c) Staggered magnetization along $x$ shows increasing staggered order with $J'$ which becomes constant close to zero field. (d) Magnetization curves remain linear for $J' > J'_{max}$, but the slope decreases for even larger $J'$. (e) $\alpha_{max}$ increases with intraladder coupling $\gamma$. (f) Interladder coupling $J'_{max}$ at which the slope is maximal decreases with increasing $\gamma$. (g) Staggered magnetization along $x$ at zero field at a function of $J'$ shows saturation and increases with $\gamma$ from $m^x_a(h=0) = 0.25$ for $\gamma = 0$.}
 \label{fig:fig3_SI_2}
\end{figure}

In fig. \ref{fig:fig3_SI_2}, we focus on the magnetization curves for a spin ladder with weak intraladder coupling $\gamma = 0.1$ and larger interladder coupling $J'$. In this case, the system is gapped, and we can use the slope of $m^z(h)$ to determine $J'$. In fig. \ref{fig:fig3_SI_2}(a), we present the magnetization close to zero field for a wide range of $J'$ values. We observe that the slope increases up to $J'_{max} \approx 1.7$, see also \ref{fig:fig3_SI_2}(b), similar to $\tilde{h}_{c_1}$. Figure \ref{fig:fig3_SI_2}(c) shows that all of these large interladder couplings $J' \geq J'_c$ lead to a staggered order at zero field. Interestingly, we observe in fig. \ref{fig:fig3_SI_2}(d) that the slope of $m^z(h)$ decreases for even larger interladder couplings $J' > J'_{max}$, while the staggered order continuously increases.

In fig. \ref{fig:fig3_SI_2}(e), we present the maximum slope $\alpha_{max}$ corresponding to the data from fig. \ref{fig:fig3b}(a). We observe that  $\alpha_{max}$ only increases by a small margin with larger $\gamma$. Figure \ref{fig:fig3b}(a) shows that the slope continues to increase monotonically up to a large $J'_{max}$, which is presented in panel (f). We see that $J'_{max}$ decreases with increasing intraladder coupling. The staggered order at zero magnetic field increases with $J'$ and eventually saturates, see \ref{fig:fig3_SI_2}(g). We know that $m^x_a(J' \rightarrow \infty) = 1/4$ for $\gamma = 0$. This limit depends on the number of coupled rung, see also fig. \ref{fig:fig_phase_diagram_gamma_0}. The saturation value of $m^x_a$ is reached at large $J'$ and can be used to determine $\gamma$ if one can vary the interladder coupling, for example, by external pressure.

\begin{figure}[htbp!]
 \centering
 \begin{adjustbox}{center}
   \includegraphics[width=1\columnwidth]{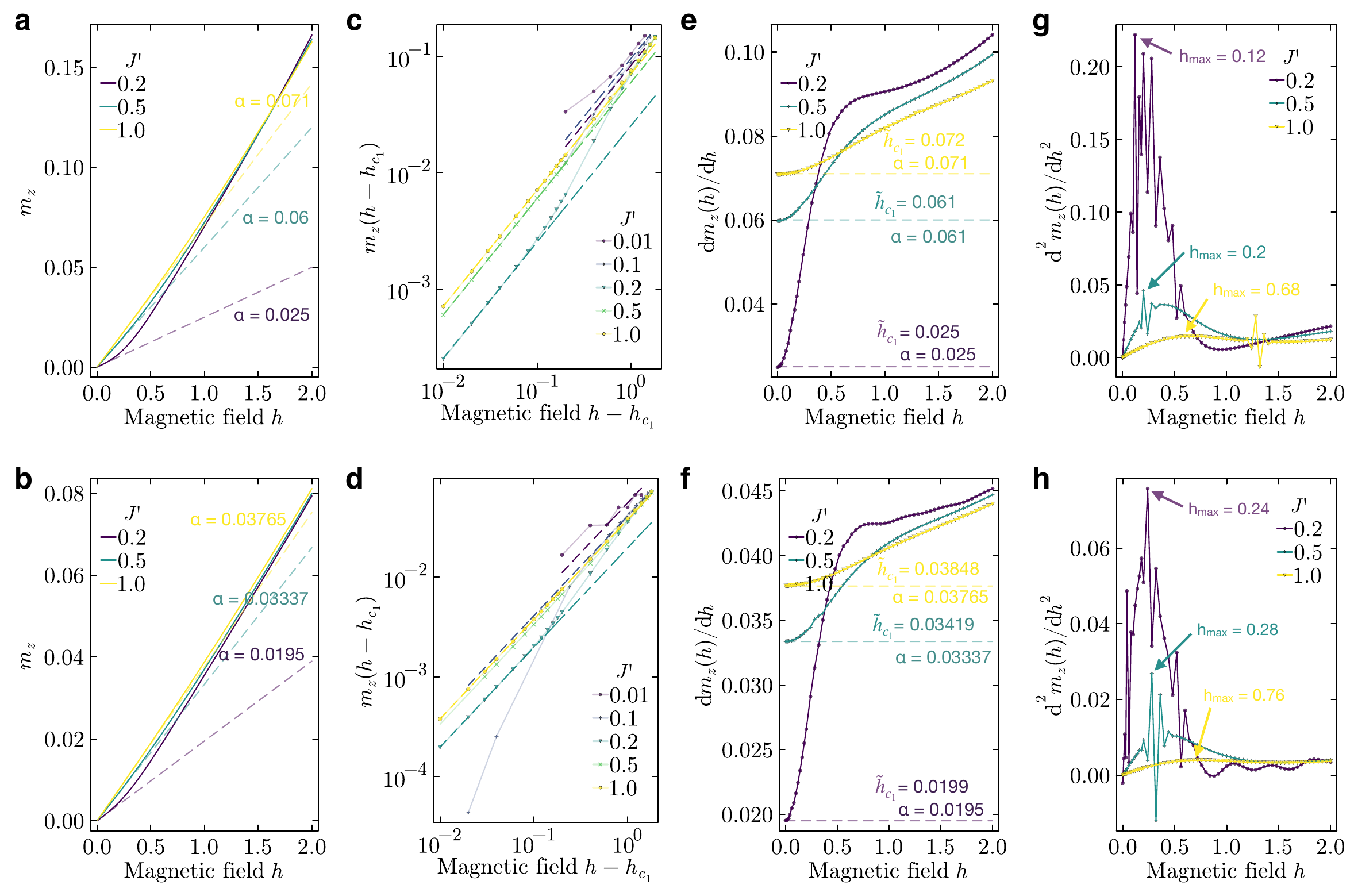}
 \end{adjustbox}
 \caption{Additional data for spin ladder with large intraladder coupling treated with mean-fields on every other rung for different magnetic fields $h$ and different interladder couplings between $J' = 0.01$ and $J' = 1$ for $\gamma = 1.5$ (top) and $\gamma = 3$ (bottom). (a) - (b) Magnetization along $z$ shows linear behavior with different slopes.  (c) - (d) Log-log plot of $m^z$ close to $h_{c_1}$ shows linear scaling. (e) - (f) Slope $\alpha$ of $m^z(h)$ is related to $\tilde{h}_{c_1}$. (g) - (h) Inflection points of $\alpha(h)$ are denoted as $h_{\text{max}}$.}
 \label{fig:fig4SI}
\end{figure}

In fig. \ref{fig:fig4SI}, we show additional data for the calculations in fig. \ref{fig:fig4} for $\gamma = 1.5$ (top row) and $\gamma =3$ (bottom row). In fig. \ref{fig:fig4SI}(a) and (b), we plot the magnetization along $z$ with a linear scaling close to zero field. We observe that the slope increases with $J'$, and similarly, the critical field $\tilde{h}_{c_1}$ at which the magnetization curve becomes nonlinear increases. Obviously, $\tilde{h}_{c_1} \approx \alpha$ for small deviations. In fig. \ref{fig:fig4SI}(e) and (f), we show that $\tilde{h}_{c_1}$ can also be related to $J'$, see eq. \ref{eq:tilde_hc1}.
We also observe that for larger intraladder coupling, the slope $\alpha$ decreases for fixed $J'$. The inflection points of $\alpha(h)$ are $\tilde{h}_{c_2}$ and be used with eq. \ref{eq:tilde_hc2} to determine a scale for $J'$.

\begin{figure}[htbp!]
 \centering
 \begin{adjustbox}{center}
   \includegraphics[width=1\columnwidth]{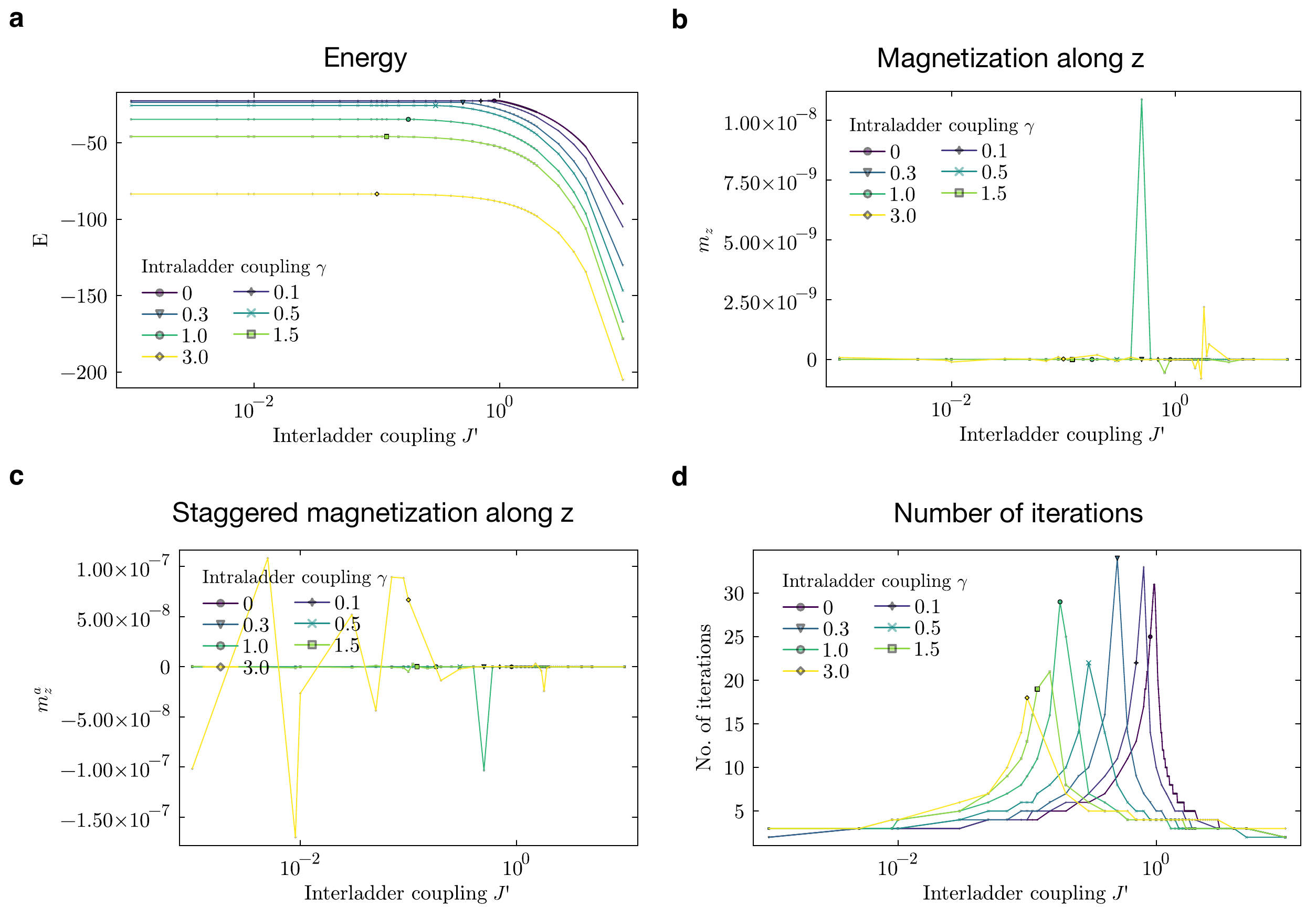}
 \end{adjustbox}
 \caption{Additional data for a single spin ladder with different intraladder couplings $\gamma$ treated with mean-fields on every second rung versus interladder coupling $J'$. All data is for zero magnetic field $h=0$. (a) Ground state energy. (b) The magnetization along $z$ is always zero without a magnetic field. (c) Staggered order along $z$ is below $10^{-7}$ for all calculations. (d) Close to $J'_c$, the number of iterations needed to reach convergence increases.}
 \label{fig:fig5SI}
\end{figure}

Figure \ref{fig:fig5SI} shows additional data for the calculations presented in fig. \ref{fig:fig5}, including the ground state energy (a) and magnetization along $z$ (b). The latter observable indicates good convergence of our DMRG calculations as the magnetic order along $z$ is below $10^{-9}$ for nearly all calculations, independent of interladder coupling $J'$. Additionally, we observe that there is also no staggered order along $z$, see \ref{fig:fig5SI}(c). The number of iterations increases close to the boundary between unpolarized order and staggered order, in agreement with fig. \ref{fig:fig_phase_diagram_SI}.

\end{document}